\documentclass[11pt,a4paper]{article}

\usepackage[english]{babel}
\usepackage{fancyhdr}
\usepackage{amsmath}
\usepackage{amssymb}
\usepackage{amsfonts}
\usepackage{psfrag}
\usepackage[applemac]{inputenc}
\usepackage{graphicx,wrapfig}
\usepackage[bf,footnotesize]{caption}
\usepackage{bm}
\usepackage[dvipsnames]{xcolor}
\usepackage[colorlinks=True, citecolor=blue, linkcolor=blue, urlcolor=blue,linktocpage]{hyperref}
\usepackage[sort&compress,numbers, merge]{natbib}
\usepackage{amsthm,bm}
\usepackage{gensymb}
\usepackage{subfig}
\usepackage{physics}
\usepackage{array}
\usepackage{tcolorbox, mathtools}
\usepackage{soul}
\tcbuselibrary{skins}
\newtcolorbox{mybox}[1][]{before=\centering, drop fuzzy shadow, enhanced, colframe=blue, fonttitle=\bfseries, title=#1, center title}
\usepackage{aas_macros}

\newcommand{\mpl}{M_\mathrm{P}}
\newcommand{\e}{\epsilon}

\newcommand{\hie}{hierarchical }

\begin{document}


\begin{titlepage}
	\begin{center}
		
		\vspace{2.0cm}
		{\large\bf  
			Effective two-body approach to the hierarchical three-body problem: quadrupole to 1PN}

		\vspace{1.0cm}
		\renewcommand{\thefootnote}{\fnsymbol{footnote}}
		{\small \bf 
			Adrien Kuntz $^{a,b}$,
			Francesco Serra $^{a,b}$, Enrico Trincherini $^{a,b}$\footnote{E-mail:
				\href{adrien.kuntz@sns.it}{adrien.kuntz@sns.it},
				\href{francesco.serra@sns.it}{francesco.serra@sns.it},	\href{enrico.trincherini@sns.it}{enrico.trincherini@sns.it}}
		}
		
		\vspace{0.7cm}
		{\it\footnotesize
			${}^a$Scuola Normale Superiore, Piazza dei Cavalieri 7, 56126, Pisa, Italy\\
			${}^b$INFN Sezione di Pisa, Largo Pontecorvo 3, 56127 Pisa, Italy\\
		}
		
		\vspace{0.9cm}
		\abstract{\noindent
		Many binary systems of interest for gravitational-wave astronomy are orbited by a third distant body, which can considerably alter their relativistic dynamics.
		Precision computations are needed to understand the interplay between relativistic corrections and three-body interactions. 
		We use an effective field theory approach to derive the effective action describing the long time-scale dynamics of hierarchical three-body systems up to 1PN quadrupole order.
		At this level of approximation, computations are complicated by the backreaction of small oscillations on orbital time-scales as well as deviations from the adiabatic approximation.
		We address these difficulties by eliminating the fast modes through the method of near-identity transformations. This allows us to compute for the first time the complete expression of the 1PN quadrupole cross-terms in generic configurations of three-body systems. We numerically integrate the resulting equations of motion and show that 1PN quadrupole terms can affect the long term dynamics of relativistic three-body systems.
		}
	\end{center}
\end{titlepage}
\renewcommand{\thefootnote}{\arabic{footnote}}
\setcounter{footnote}{0}

\tableofcontents
\newpage

\section{Introduction}

\begin{figure}
\centering
\includegraphics[width=0.3\columnwidth]{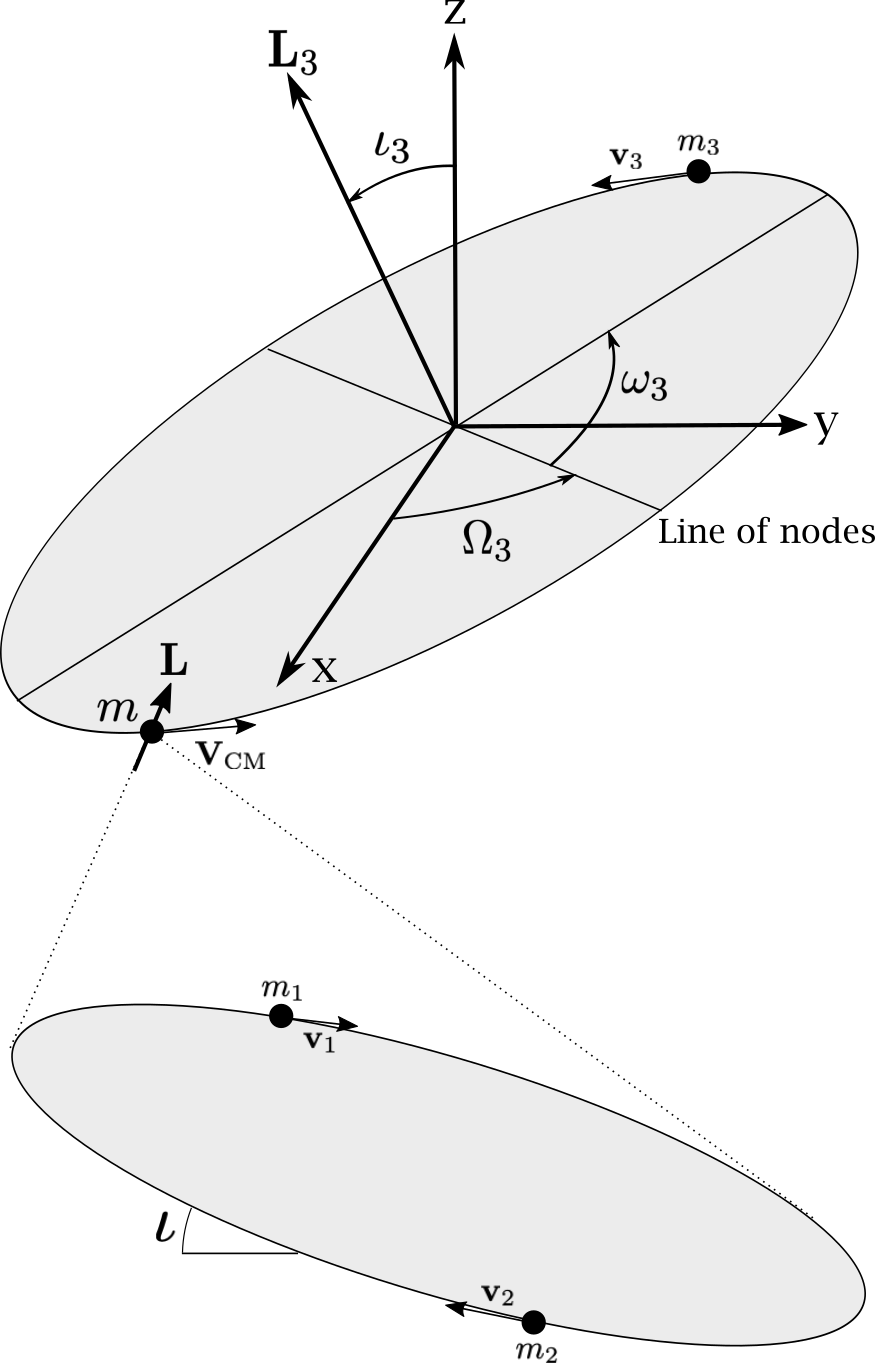}
\caption{Illustration of the "effective two-body" description and of osculating elements. The inner binary is replaced with a point-particle whose spin and multipole moments are related to the osculating elements of the inner orbit. }
\label{fig:effective_two_body}
\end{figure}

The evolution of binary systems can be significantly affected by relativistic effects as well as by the presence of a third celestial body. The interplay between these effects is especially relevant in view of recent observations of triple systems with pulsars~\cite{Ransom_2014} as well as future detections of relativistic three-body systems expected with the gravitational-wave (GW) spatial interferometer LISA~\cite{Robson_2018, Inayoshi:2017hgw}. Three-body effects have been studied to understand the merger rate of compact objects~\cite{10.1093/mnras/stu039, Seto:2013wwa, Stephan_2016, PhysRevD.84.044024, Hoang_2018, Naoz_2013, 2020ApJ...901..125D}, the transits of exoplanets~\cite{Martin_2015, 10.1093/mnrasl/slv139, Mazeh_1997} or the evolution of triple star systems~\cite{1979A&A....77..145M, Liu_2014, 2015MNRAS.451.1341L, 2014ApJ...785..116L}. Triple systems often come in a \hie setting where one of the constituents is far away from the other two. 
These configurations are conveniently described in terms of an "inner binary" elliptic motion composed of the two closest objects, and an "outer binary" elliptic motion made of the inner binary itself and the outer object, as depicted schematically in Fig~\ref{fig:effective_two_body}. The long-timescale evolution of these orbits presents many interesting features already at the Newtonian level~\cite{Naoz_2016}, the most well-known example being the Kozai-Lidov (KL) oscillations of the eccentricity of the inner binary~\cite{1962AJ.....67..591K, LIDOV1962719}. These effects are usually studied through the method of double averaging \cite{Naoz_2016}, in which quick oscillations on time-scales of order of the orbital periods are eliminated from the dynamics, leaving only information about evolution on timescales larger than the orbital periods. This method is known to work whenever the system is away from resonances; the latter make the whole analysis much less straightforward, see e.g. \cite{Kuntz:2021hhm}.

When relativistic corrections to the Newtonian dynamics are considered, in the so-called Post-Newtonian (PN) expansion, the long time-scale dynamics undergoes radical changes, even when the system remains into a hierarchical configuration. For instance, the periastron precession can suppress high eccentricity oscillations when the precession timescale is much shorter than the timescale characterizing KL oscillations~\cite{Blaes_2002, 10.1111/j.1365-2966.2007.11694.x, 10.1093/mnras/stu039, Biscani:2013zva, Ford_2000}. Even if the system can be described by the inner and the outer orbit, studying the effects of relativistic corrections on its evolution over long time-scales remains in general quite difficult especially due to effects at higher-orders in the PN expansion~\cite{Naoz_2013}, which will usually leave an imprint on the dynamics after a parametrically large time. 
It is thus important to compute three-body relativistic interactions to a high accuracy in order to evolve these systems on long timescales. 
One option is to approach the problem with a numerical relativistic three-body solver~\cite{PhysRevD.83.084013, PhysRevD.84.104038, 10.1093/mnras/stw1590,Lousto:2007ji,Gupta:2019unn}, however this method is heavily time-consuming. On the other hand, analytic methods are sparse at best~\cite{ Biscani:2013zva, Ford_2000,Naoz_2013,PhysRevD.89.044043,PhysRevLett.120.191101,Will:2014wsa, Lim:2020cvm}.

Motivated by these developments and by the ensuing difficulties, we recently introduced a new Effective Field Theory (EFT) approach to the relativistic, \hie three-body problem~\cite{Kuntz2021}. The central idea of this method is to take advantage of the two small parameters characterizing hierarchical three-body systems: the typical velocity of the inner bodies divided by light speed, $v$  ($ c=1 $), and the ratio $\varepsilon = a/a_3$ between the semimajor axis of the inner binary $a$ and the one of the outer binary $a_3$.
Thanks to the double perturbative expansion with respect to these parameters, it is possible to match the dynamics of a \hie three-body system to a simpler two-body interaction, in which the inner binary is described as a single point-particle endowed with multipole moments. This is achieved by employing the EFT techniques developed for the relativistic two-body problem~\cite{porto_effective_2016, goldberger_effective_2006} and by performing the averaging procedure at the level of the Lagrangian. Most noticeably, the EFT approach exploits symmetries that are manifest in the effective Lagrangian, restricting the form of the allowed interaction terms. Moreover, working with a single functional rather than with several equations of motion makes it simpler to setup a systematic study of the three body system.

In \cite{Kuntz2021} we presented the EFT setup and derived the effective Lagrangian describing the system on long time-scales up to 1PN dipole order, i.e. up to order $v^2 \varepsilon^{3/2}$ beyond the leading Newtonian interaction. Instead, in the present work we extend this computation up to 1PN quadrupolar order, i.e. $v^2 \varepsilon^{5/2}$ beyond leading order.
At this order, computations are substantially more complex with respect to our previous study \cite{Kuntz2021}. 
A first source of complexity is due to the averaging procedure. At lower orders the averaging can be performed in the so-called adiabatic approximation, i.e. neglecting variations of slowly evolving variables during the average over the period of both orbits. Instead, when accounting for terms of mixed quadrupolar and 1PN order, deviations from adiabaticity must be taken into account. In addition to this, backreaction from quickly oscillating terms that are suppressed in amplitude will also affect the averaging, contrarily to what happened at lower orders. We address these complications by following the method of near-identity transformations \cite{Murdock}, which allows to consistently implement the averaging procedure to any order of accuracy.
While several authors already studied quadrupolar couplings at 1PN order~\cite{Fang:2019mui, PhysRevD.102.104002, PhysRevD.89.044043, PhysRevLett.120.191101, 2019ApJ...883L...7L, Lim:2020cvm, Naoz_2013,Migaszewski:2008tp}, we are aware of only three which took into account these deviations from the adiabatic approximation~\cite{PhysRevD.89.044043,PhysRevLett.120.191101,Will:2014wsa, Lim:2020cvm}. However, we believe that we give in this work the first complete expressions of quadrupolar 1PN terms. Indeed, {only the particular case of a circular outer orbit is considered in}~\cite{PhysRevD.89.044043,PhysRevLett.120.191101,Will:2014wsa} {neglecting some} PN interaction{s} that we describe in this work. On the other hand, the {derivation in}~\cite{Lim:2020cvm} reports a puzzling result that we will mention in Section~\ref{sec:double_averaged_Lagrangian}.

Another source of complexity lies in finding the suitable dynamical variables to efficiently package relativistic corrections in our results. While the idea at the core of our approach of identifying the inner binary to a spinning point-particle with multipole moments is very intuitive, providing a definite relation between the variables describing the inner binary and the parameters of the effective point-particle is subtle in practice. For example, in our previous work~\cite{Kuntz2021} we showed how the choice of a Spin Supplementary Condition (a gauge condition on the spin tensor of the effective point-particle~\cite{HANSON1974498, Levi:2015msa}) is related to the center-of-mass choice of the inner binary. In the present computations, two new similar subtleties arise. The first one concerns the definition of osculating elements describing both inner and outer orbits. In our previous work, we followed the usual convention and used the osculating \textit{orbital} elements defined as the parameters of the ellipse instantaneously tangent to the trajectory (described with positions and velocities). However, at 1PN quadrupolar order we find that it becomes more convenient to use osculating \textit{contact} elements, which are defined through momenta rather than velocities~\cite{relativistic_celestial_mechanics}. Since PN corrections induce a non-trivial relation between momentum and velocity, these two sets of osculating elements will differ in general. We will give in Section~\ref{sec:contact} the precise definition of contact elements, and we will elaborate more on their difference with respect to orbital elements in Appendix~\ref{app:contact}. One remarkable conclusion of the present analysis is that, while the slowly evolving part of the contact semimajor axes are conserved throughout the evolution of the system (as is common in long-timescale evolution of triple systems~\cite{Naoz_2016}), their orbital counterpart features small variations over long time-scales, which offers a new point of view on earlier findings of~\cite{PhysRevD.89.044043,PhysRevLett.120.191101,Will:2014wsa, Lim:2020cvm}. 
	
The second subtle point in the matching between inner binary and point-particle is that the quantities describing the inner binary system are inherently defined in the rest frame of its center of mass, which is accelerating because of the presence of the third body. This entails non-trivial relations between the absolute positions of the inner binary components, defined in the rest frame of the three-body center-of-mass, and the relative quantities defined in the binary rest frame, as we show in Section~\ref{sec:CM}. As far as we know, this point went so far unnoticed in the relativistic three-body literature. While this step just amounts to a redefinition of the osculating elements of the inner binary, it proves to be crucial in order to perform correctly the matching procedure described in Section~\ref{sec:matching}. 

Let us now describe in more detail the organization of this article. We will begin by summarising our "effective two-body" approach in Section~\ref{sec:summary}, which is defined by four main steps. Steps 1 and 2 will then be performed in Section~\ref{sec:average}, where the core of our matching procedure is explained. In order to keep the discussion as simple as possible, we have deferred the computation of beyond-adiabatic corrections to Appendix~\ref{app:beyond_adiab} and use only the final result of this appendix in the main text. We then perform the two final steps of our approach in Section~\ref{sec:double_averaged_Lagrangian}, where we integrate out the gravitational field due to the outer object. Finally, in Section~\ref{sec:numerical} we provide a numerical solution implementing the new relativistic interaction derived in the present work and we show how it influences the long time-scale dynamics in the case of a particular three-body system. The rest of the Appendices are devoted to: a presentation of the averaging procedure that we employ (Appendix~\ref{app:toymodel}); a discussion of the conservation of the semimajor axis (Appendix~\ref{app:semimajor}); a review of the Lagrange Planetary Equations (Appendix~\ref{app:LPE}); the derivation of the expressions connecting the absolute coordinates of the two inner bodies in the three body rest frame to their relative coordinates in the inner binary rest frame (Appendix~\ref{app:CM}); an independent computation of the so-called quadrupole-squared terms of~\cite{Will:2020tri} (Appendix~\ref{app:quad_squared}); and the relation between contact and orbital elements (Appendix~\ref{app:contact}).

\subsection{Dictionary of symbols}

{As a preliminary, we define the quantities that characterize the three body problem and list the symbols that we are going to use through the work. We will work out the results in terms of the contact elements, the variables in which the PN Hamiltonian has the simplest expression. With respect to the usual Newtonian orbital elements, contact elements include PN corrections. We review their precise definition in Section \ref{sec:summary} and examine their difference from orbital elements in Appendix \ref{app:contact}.}
\begin{itemize}
\item $\bm y_1$, $\bm v_1$, $\bm y_2$, $\bm v_2\,$: positions and velocities of the two constituents of the inner orbit, of masses $m_1$ and $m_2$;
\item $\bm y_3$, $\bm v_3\,$: position and velocity of the external perturber, of mass $m_3$;
\item $\bm Y_\mathrm{CM}$, $\bm V_\mathrm{CM}\,$: position and velocity of the center-of-mass of the inner binary, defined in Eq~\eqref{eq:CM_1PN};
\item $\bm r$, $\bm v\,$: relative variables in the inner binary instantaneous rest frame, defined in Section~\ref{sec:CM}. Note that $\bm r = \bm y_1-\bm y_2$ and $\bm v = \bm v_1-\bm v_2$ only at lowest PN order, as explained in Section~\ref{sec:CM};
\item {$ \bm p_A $ ($ A=1,2,3 $), $ \bm P_{\mathrm{CM}} $, $ \bm p $: conjugate momenta to the positions of the three bodies, position of the center of mass, and relative separation between two inner bodies;}
\item $r = \vert \mathbf{r} \vert$, $\mathbf{n} = \mathbf{r}/r$, $\bm R = \bm Y_\mathrm{CM} - y_3$, $R = \vert \mathbf{R} \vert$, $\mathbf{N} = \mathbf{R}/R$, $\bm V = \bm V_\mathrm{CM} - \bm v_3$;
\item $m=m_1 +m_2$ is the mass of the inner binary, $\mu = m_1m_2/m$ is the reduced mass of the inner {binary}, $X_1 = m_1/m$, $X_2=m_2/m$ and $\nu = \mu/m$ are its {characteristic} mass ratios;
\item $a$ [$a_3$]: {contact} semimajor axis of the inner [outer] orbit;
\item $e$ [$e_3$]: {contact} eccentricity of the inner [outer] orbit;
\item $ \mathcal{E} = m - G_N m \mu/(2a) $: total (relativistic) energy of the inner binary;
\item $M=\mathcal{E}+m_3$ is the total mass of the effective two-body system. Similarly, $\mu_3 = m_3 \mathcal{E}/M$ is its reduced mass, $X_3 = m_3/M$, $X_\mathrm{CM}=\mathcal{E}/M$ and $\nu_3 = \mu_3/M$ are the mass ratios characterizing the outer {orbit};
\item $n=\sqrt{G_N m/a^3}$ [$n_3 = \sqrt{G_N M/a_3^3}\,$]: {contact} frequency of the inner [outer] orbit;
\item $\bm{ \alpha}$, $\bm{ \beta}$, $\bm{ \gamma}$ [ $\bm{ \alpha}_3$, $\bm{ \beta}_3$, $\bm{ \gamma}_3$]: {contact} orthonormal basis of vectors characterizing the inner [outer] orbit, aligned respectively along the {contact} semimajor axis (pointing towards the pericenter), the semiminor axis, and the angular momentum;
\item $\Omega$, $\omega$, $\iota$ [$\Omega_3$, $\omega_3$, $\iota_3$]: {contact} angles characterizing the orientation of the inner [outer] orbit, defined by $ \bm{\alpha} = R_z(\Omega) R_x(\iota) R_z(\omega)  \mathbf{u}_x$ where the $R_{x_i}$'s are rotation matrices along the given axis $ x_i $, and similarly for the other vectors;
\item $u$ and $\eta$ [$u_3$ and $\eta_3$]: {contact} mean and eccentric anomaly of the inner [outer] orbit. They are defined by $\eta - e \sin \eta = u$ and $u = n t+\sigma$ [$\eta_3 - e_3 \sin \eta_3 = u_3$ and $u_3 = n_3 t+\sigma_3$] where $\sigma$ and $\sigma_3$ are the {contact} mean anomalies at initial time;
\item $L$, $G$, $H$ [$L_3$, $G_3$, $H_3$]: {contact }conjugate momenta to $u$, $\omega$ and $\Omega$ respectively [$u_3$, $\omega_3$ and $\Omega_3$ respectively], defined in Eq.~\eqref{eq:def_L_G_H};
\item $\bm{J} = \mu \sqrt{G_N ma(1-e^2)} \bm{\gamma}$ [$\bm{J}_3 = \mu_3 \sqrt{G_N M a_3(1-e_3^2)} \bm{\gamma}\,$]: {1PN corrected (i.e. contact)} Newtonian angular momentum vector of the inner [outer] orbit.
\end{itemize}
{In addition to these symbols, we will use tilded contact elements to indicate orbital elements, which will only appear in Appendix \ref{app:contact}.} We use the mostly positive metric signature and set $c=1$. Greek letters refer to spacetime indices ranging from 0 to 3, while latin indices are spatial and range from 1 to 3. $\epsilon_{\alpha \beta \gamma \delta}$ denotes the totally antisymmetric Levi-Civita tensor, not to be confused with {our \hie expansion parameter} $\varepsilon \equiv a/a_3$. {The symmetric and anti-symmetric parts of a tensor are defined with parenthesis and brackets respectively, $A_{(\mu \nu)} = (A_{\mu \nu} + A_{\nu \mu})/2$ and $A_{[\mu \nu]} = (A_{\mu \nu} - A_{\nu \mu})/2$.}

\section{The EFT approach to the relativistic hierarchical three body problem} \label{sec:summary}

We consider a \hie system in which the inner binary distance vector $r$ is much smaller than its outer binary counterpart $R$. The idea at the core of \cite{Kuntz2021} is that one can establish a correspondence between the Lagrangian of a \hie three-body system and an effective Lagrangian describing a two-body system, where one of the two bodies is a spinning point-particle endowed with multipole moments representing the inner binary. This identification was performed up to dipolar order in \cite{Kuntz2021}. Here we summarize briefly the main steps of the computation:
\begin{enumerate}
\item We start with the GR action describing two bodies (the inner system) in the presence of a gravitational field:
\begin{equation}
S = \frac{\mpl^2}{2} \int \mathrm{d}^4 x \sqrt{-g} \; R - \sum_{A=1,2} m_A \int \mathrm{d}t \sqrt{- g_{\mu \nu} v_A^\mu v_A^\nu} \; ,
\end{equation}
where $v_A^\mu = (1, \mathbf{v}_A)$ is the coordinate velocity of the point-particle. The line element of our spacetime is decomposed in the non-relativistic {limit} following the space+time splitting presented in~\cite{Kol:2007bc, Kol:2010si}:
\begin{equation}
\mathrm{d}s^2 = - e^{2 \phi} \left(\mathrm{d}t - A_i \mathrm{d}x^i \right)^2 + e^{-2 \phi} \gamma_{ij} \mathrm{d}x^i \mathrm{d}x^j \; ,
\end{equation}
where it is sufficient to consider the spatial part of the metric as diagonal for our 1PN computations: $ \gamma_{ij} \simeq\delta_{ij}$ . The two metric fields are decomposed using the background field method as $\phi = \bar \phi + \tilde \phi$, $A_i = \bar A_i + \tilde A_i$, where  the tilde quantities correspond to an external arbitrary field (which will be related to the gravitational field of the third distant object in Step 3), while we integrate out the barred quantities corresponding to gravitons exchanges between the two bodies. At 1PN order, the resulting effective Lagrangian is
\begin{align}\label{eq:L1PN_unexpanded}
\begin{split}
\mathcal{L}  = \mathcal{L}_\mathrm{EIH}  - m_1 \tilde \phi(\mathbf{y}_1) \left( 1 + \frac{3}{2} v_1^2 \right)
- \frac{m_1}{2} \tilde \phi(\mathbf{y}_1)^2 + m_1 \tilde{\mathbf{A}}(\mathbf{y}_1) \cdot \mathbf{v}_1 + \frac{G_N m_1 m_2}{r} \tilde \phi(\mathbf{y}_1)
 + (1 \leftrightarrow 2) \; ,
\end{split}
\end{align}
where $\mathcal{L}_\mathrm{EIH}$ is the Einstein-Infeld-Hoffmann Lagrangian given by \cite{einstein_gravitational_1938}:
\begin{align}\label{eq:LEIH}
\begin{split}
\mathcal{L}_\mathrm{EIH} &=  \frac{1}{2} m_1 v_1^2 + \frac{1}{2} m_2 v_2^2 + \frac{G_N m_1 m_2}{r} \\ &+ \frac{1}{8} m_1 v_1^4 + \frac{1}{8} m_2 v_2^4 + \frac{G_N m_1 m_2}{2 r} \bigg[ 3 v_1^2 + 3 v_2^2  - 7 \mathbf{v}_1 \cdot \mathbf{v}_2 - \mathbf{v}_1 \cdot \mathbf{n} \; \mathbf{v}_2 \cdot \mathbf{n} - \frac{G_N m}{r} \bigg] \; .
\end{split}
\end{align}
\item The Lagrangian~\eqref{eq:L1PN_unexpanded} is expanded in multipoles around the center-of-mass of the inner orbit, then averaged over one period of the inner orbit in order to describe the dynamics on longer timescales.{ As discussed in Section \ref{sec:average}, in this way we obtain the classical effective Lagrangian for the long-timescale modes of the system.} This Lagrangian describes a composite spinning point-particle coupled to an arbitrary external gravitational field. The multipole moments of the composite object can be described in terms of either its osculating elements or its contact elements, {see section \ref{sec:contact}}. For instance the orbital elements would be defined through the position and velocity of the inner binary by
\begin{align}
\begin{split} \label{eq:DefOsculatingElements}
\tilde{a} &= - \frac{G_Nm}{2} \left( \frac{v^2}{2} - \frac{G_N m}{r} \right)^{-1} \; , \\ 
 \tilde{e} \bm {\tilde{\alpha}} &= \frac{1}{G_N m} \mathbf{v} \cross \left( \mathbf{r} \cross \mathbf{v} \right) - \frac{\mathbf{r}}{r} \; , \\
  \bm{\tilde{\gamma}} &= \frac{\mathbf{r} \cross \mathbf{v}}{\sqrt{G_N m\tilde{a}(1-\tilde{e}^2)}} \; .
\end{split}
\end{align}
In this equation, $\tilde{a}$ and $\tilde{e}$ are the semimajor axis and eccentricity of the orbit, while $\bm{ \tilde{\gamma}}$ and $\bm{\tilde{ \alpha}}$ are two unit vectors respectively directed along the Newtonian angular momentum and the Newtonian perihelion of the inner orbit. In this work, in order to handle higher order corrections in a compact way, we find convenient to use contact elements, which we indicate with untilded symbols and define below, see Eq. \eqref{eq:DefContactElements}.

\item We add the action of a third point-particle and, as in the first step, integrate the external fields $\tilde \phi$, $\tilde A_i$ mediating interactions between the third point-particle and the effective spinning point-particle representing the inner binary. The number of terms to take into account is dictated by a set of power-counting rules generalizing the NRGR approach~\cite{goldberger_effective_2006, porto_effective_2016} to the \hie three-body problem, and described in \cite{Kuntz2021}. The two dimensionless parameters controlling the perturbative expansion are the usual PN parameter $v^2 = G_N m/a$ and the ratio of semimajor axes $\varepsilon = a/a_3$, where $a_3$ is the semimajor axis of the outer orbit.

\item We finally average the Lagrangian over the outer binary timescale. At this final step, the Lagrangian depends only on the osculating elements of both inner and outer orbit. Note that the osculating elements of the outer orbit are defined similarly as in~\eqref{eq:DefOsculatingElements}, replacing all inner quantities with outer ones. In particular, at first order in the PN expansion, the total mass of the outer binary will be $m_3 + \mathcal{E}$, where $\mathcal{E} = m - G_N m / (2a)$ is the total energy of the inner binary to Newtonian order. Thus, the osculating elements of the outer orbit implicitly contain post-Newtonian corrections, which is a difference of our approach compared to previous ones (e.g.~\cite{PhysRevD.89.044043, Lim:2020cvm, Naoz_2013}) which use the total mass $m+m_3$ to define the outer orbit osculating elements.
\end{enumerate}

After these four steps, we obtain a Lagrangian describing the interactions between the inner and outer orbits up to order $v^2 \varepsilon^{5/2}$ on timescales much larger than the orbital periods:

\begin{mybox}
\begin{equation} \label{eq:totLagrangian}
\mathcal{L} = \bm J \cdot \bm \Omega +  \bm J_3 \cdot \bm \Omega_3 + 3 \mu \frac{G_N^2 m^2}{a^2 \sqrt{1-e^2}}  + 3 \mu_3 \frac{G_N^2 M^2}{a_3^2 \sqrt{1-e_3^2}} - \frac{4m + 3 m_3}{2 m}  \Omega_\mathrm{prec} \bm J \cdot \bm J_3 + \left\langle\mathcal{L}_\mathrm{quad}^{\leq v^2 \varepsilon^{5/2}}\right\rangle \; .
\end{equation}
\end{mybox}
In this equation, $\mathbf{J}$ ($\mathbf{J}_3$) and $\bm \Omega$ ($\bm \Omega_3$) are the angular momentum and rotation vectors of the inner (resp. outer) orbits, defined by
\begin{equation}\label{eq:def_J_Omega}
\bm{J} = \mu \sqrt{G_N ma(1-e^2)} \bm{\gamma} \; , \quad \bm{\Omega} =  \bm{\alpha} \times \dot{\bm{ \alpha}} \; ,
\end{equation}
with analogous formulas for the outer orbit. Let us comment on each component of Eq~\eqref{eq:totLagrangian}. The first two terms are the spin kinetic terms of the two orbits. Once a variational principle is applied, they will give rise to first-order evolution equations for the planetary elements of the two orbits, named Lagrange Planetary Equations (LPE), see Appendix~\ref{app:LPE}.
Note that, as mentioned before, the spin $\mathbf{J}_3$ of the outer orbit is defined by using the 1PN energy $\mathcal{E}$ of the binary system as its effective mass, so that the spin kinetic term of the outer orbit secretly hides post-Newtonian terms (the other term where $\mathbf{J}_3$ appears in \eqref{eq:totLagrangian} is already of 1PN order, so that in this term the difference between using $m$ or $\mathcal{E}$ is of 2PN order).
The two next terms correspond to the well-known 1PN potentials inducing perihelion precession of the two orbits at 1PN order~\cite{Misner:1973prb}. We {refer to them as} {internal} Lagrangians since they would be present even without any interaction between the two orbits.
The third term is a coupling between the angular {momenta} of the two orbits at dipole order $v^2 \varepsilon^{3/2}$. This term, whose effect on the orbits of \hie systems was already studied e.g{.} in~\cite{PhysRevD.102.104002, Fang:2019mui, 2019ApJ...883L...7L}, was the highest-order coupling in the $\varepsilon$ expansion which we described in \cite{Kuntz2021}. It involves a precession frequency given by
\begin{equation} \label{eq:omegaPrec}
 \Omega_\mathrm{prec} = \frac{G_N }{a_3^3 (1-e_3^2)^{3/2}} \; .
\end{equation}
Finally, the term $ \mathcal{L}_{\mathrm{quad}}^{\leq v^2\varepsilon^{5/2}} $ encodes the contributions from quadrupole-suppressed 0PN and 1PN interactions to the long time-scale  dynamics, which we will compute in the next sections and which is the main result of the present work.

\subsection{Contact elements}\label{sec:contact}
The quantities above are given in terms of contact elements, which differ from the usual Newtonian orbital elements by PN corrections. Such differences will only become relevant if one considers high enough terms in the perturbative analysis of the problem, which happens to be the case in the present work. The key difference is that contact elements are defined in terms of the canonical momenta of the system, rather than in terms of the velocities (see e.g.~\cite{relativistic_celestial_mechanics}). This allows to keep track of some PN corrections in a compact way, since the conjugate momenta will receive corrections as soon as the interaction Lagrangian depends on the velocities. 

Let us now define the contact elements. In the rest frame of the inner binary center of mass, we can write the Lagrangian of the inner binary system as
\begin{align}
	\mathcal{L}=\dfrac{1}{2} \mu v^2+\dfrac{G_N m\mu}{r}+\mu \mathcal{R}\;,
\end{align}
where $\mathcal{R}$ encodes any term in the Lagrangian beyond Newton's expression for the inner binary system.
The conjugate momentum to the coordinate $ \bm r $ will be:
\begin{align}
	\bm p =\mu\bm v+ \mu \partial\mathcal{R}/\partial \bm v\;.
\end{align}
Then, rather than using Eq. \eqref{eq:DefOsculatingElements}, we will define the contact elements as follows:
\begin{align}
	\begin{split} \label{eq:DefContactElements}
		a &= - \frac{G_Nm}{2} \left( \frac{p^2}{2\mu^2} - \frac{G_N m}{r} \right)^{-1} \; , \\ 
		e \bm{{\alpha}} &= \frac{1}{G_N m\mu^2} \mathbf{p} \cross \left( \mathbf{r} \cross \mathbf{p} \right) - \frac{\mathbf{r}}{r} \; , \\
		\bm{{\gamma}} &= \frac{\mathbf{r} \cross \mathbf{p}}{\sqrt{G_N m \mu^2 a(1- e^2)}} \; , 
	\end{split}
\end{align}
where the unit vector $ \bm{\gamma}$ contains two angles and it is orthogonal to $ \bm \alpha $. 
With these definitions, in analogy with the Kepler problem, we have:
\begin{align}
	\begin{split} \label{eq:ParamContactElements}
		\mathbf{r} &=  a \left( (\cos \eta -  e) \; {\bm{\alpha}} + \sqrt{1 -  e^2} \sin \eta \; { \bm{\beta}}  \right) \; , \\
		\mathbf{p} &=\sqrt{\frac{G_N m}{a}} \frac{\mu}{1 -  e \cos  \eta} \left( - \sin  \eta \; {\bm{\alpha}} + \sqrt{1 -  e^2} \cos  \eta \; { \bm{\beta}}  \right) \; .
	\end{split}
\end{align}
where $ \bm{\beta} = \bm{\gamma} \cross \bm{\alpha} $ and $\eta$ is the (contact) eccentric anomaly, defined at a time $t$ by
\begin{equation} \label{eq:def_eccentric_anomaly}
	u \equiv nt + \phi = \sqrt{\frac{G_N m}{a^3}} t + \phi = \eta - e \sin \eta \; ,
\end{equation}
with $\phi$ the initial phase, and $u$ the contact equivalent of the mean anomaly.
In the following, it will be useful to have an explicit relation between the contact elements $(a, e, u, \omega, \Omega, \iota)$ and the orbital elements $(\tilde a, \tilde e, \tilde u, \tilde \omega, \tilde \Omega, \tilde \iota)$. We derive such a relation in Appendix~\ref{app:contact}, see Eqs.~\eqref{eq:deltaa}-\eqref{eq:deltaiota}. An important outcome of Appendix~\ref{app:contact} which we will use in the main text is that the difference between orbital and contact elements is small in the sense that at \textit{any time}, they differ by a 1PN quantity (i.e. this difference cannot grow on long timescales).

\section{The point-particle EFT to quadrupolar order}
\label{sec:average}
We now derive an effective Lagrangian describing the inner binary coupled to an external gravitational field on time-scales much longer than its orbital period. We start by describing the procedure to integrate out the fast (orbital) modes, the so-called averaging, which we present in detail in Appendix \ref{app:toymodel}. We then discuss in which reference frame to define the contact elements of the inner binary, highlighting the rest frame of the inner binary's center of mass as the most suitable choice. Having done that, we present the results of the averaging procedure, which is carried out in some detail in Appendix \ref{app:beyond_adiab}. We close the section by matching the result to a world-line action for the inner binary.

\subsection{Integrating out fast modes}
In field theoretic terms, we wish to derive the effective action for the slow modes (the long time-scale dynamics) by integrating out the fast modes (the orbital dynamics). In the classical limit, this corresponds to substituting the solutions to the equations of motion for the fast modes in terms of the slow modes in the Lagrangian. On top of that, we can average this effective Lagrangian, so as to remove any quickly oscillating terms, which only carry information about the already solved short time-scale dynamics.
In our specific case, fast and slow modes are packed together in our variables, the osculating (orbital or contact) elements. Therefore, we need a way to split the dynamics over short and long time-scales. Intuitively, a method to achieve this is considering the average and the average-free part (with respect to the orbital time-scale) of the original equations of motion.
The splitting allows to solve for the fast modes in terms of the slow ones, by solving the average-free equations. This method is broadly referred to as averaging, see for instance \cite{Murdock,Sanders}. 

Despite its intuitiveness, averaging presents a few subtleties which must be clarified in order to set up a systematic and consistent procedure. 
For instance, if the period of the fast oscillations of our system changes slowly, we have to understand how to account for variations of the period in our averages. Even more, we need to account for the small changes of the slow variables over the period of a quick oscillation. Up to which order can we compute averages while keeping fixed the slow variables?

While it is easy to estimate the size of these possible corrections, the task of choosing a set-up that makes transparent how to deal with these issues is much less straightforward. For instance, one can consider whether to average with respect to time or with respect to a dynamical variable (e.g. a time-dependent angle). We can promptly see that these two choices will lead to different quantities. For instance, suppose the dynamics has periodicity with respect to an angle $ u $. If we call $ T[\ell] $ the period as a function of the slow variables $ \ell $, the average of a quantity $ A(\ell, u) $ can be defined in two different ways:
\begin{align}
	\langle A\rangle_u=&\int_{u_0}^{u_0+2\pi}\dfrac{du}{2\pi}A(\ell,u)\quad,\\ \langle A\rangle_t=&
	\dfrac{1}{T[\ell(t)]}\int^{t+T[\ell(t)]}_{t}\!\!\!\!\!\!\!\!\!\!\!\!\!\!\! dt'A(\ell,u(t'))=\dfrac{1}{T[\ell(t)]}\int_{u(t)}^{u(t)+2\pi}{du}\dfrac{dt}{du}A(\ell,u)\;.
\end{align}
Even when in the first integral $ u_0=u(t) $, these two quantities will be different as long as $ dt/du $ depends on $ u $.

Intuition suggests that the difference between the choices that one can take at the level of the setup might be akin to the difference between the various choices of renormalization scheme. While it seems reasonable that different averaging methods and choices can be followed consistently to describe the same dynamics, some of the choices that we have to make might depend dramatically on the nature of the system itself. An extreme example is the case of the so-called \textit{crude averaging}, see e.g. \cite{Sanders}, which shows that in some cases there can be problems in the convergence of the averaged solutions unless one chooses to perform the average after having rewritten the equations of motion in a certain canonical form.

In light of these possible complications, following closely \cite{Murdock} we set up our averaging procedure by means of the so called near-identity transformations for an angle-periodic system written in its canonical form. This method consists in performing a change of variables such that the system of equations becomes independent on the angle. We review this construction in Appendix \ref{app:toymodel}.
 
The choice of this method has a few relevant consequences. First, it implies that the averages (both at the level of the equations and at the level of the Lagrangian), must be taken with respect to the slowly evolving part of the mean anomaly. For practical reasons, these averages will then be expressed as integrals over the eccentric anomaly (or rather, the corresponding contact element). This makes clear that the possibly changing value of the time-period of the orbit does not lead to any corrections to the averages. Moreover, as we show in Appendix \ref{app:toymodel}, the construction of near identity transformations is such that the averages are performed, to any order, keeping the slow variables fixed in the integrand. This might seem counter-intuitive, but stems from the fact that the averages appear as a by-product of a certain transformation of our variables, rather than as a direct coarse graining of the dynamics.

Having defined the averaging procedure, our practical task is to apply it to the relativistic, hierarchical three-body problem, up to quadrupolar order. In doing so, we only have difficulties due to the somewhat large number of variables (the contact elements) and to the presence of two small parameters that characterize the perturbation theory: the ratio of semimajor axes and the typical velocity of the bodies. On top of this, as we have already remarked, we will have to perform two averaging procedures.

\subsection{Center-of-mass and relative coordinates in boosted frame} \label{sec:CM}
In order to carry out the multipole expansion, we express the two inner bodies' coordinates $\bm y_1$, $\bm y_2$ as functions of the position of the inner binary center of mass and of the relative distance between the two inner bodies, $\bm Y_\mathrm{CM}$ and $\bm r$. Then, introducing the contact elements, we can describe the inner binary as a spinning particle endowed with multipole moments, coupled to gravity. 

When this is done to 1PN order, in general one needs to account for the difference between the rest frame of the three-body center of mass and the rest frame of the center of mass of the inner binary, as illustrated in Figure~\ref{fig:change_ref}. In fact, one might express the Lagrangian in terms of contact elements that are defined in either of the two reference frames, by means of Eq. \eqref{eq:ParamContactElements}. These two frames are connected by a boost plus a translation, a transformation that is non-trivial starting at 1PN order. This entails a difference between the two sets of contact elements that one can define. Crucially, while it is natural to express the effective action in the rest frame of the three-body center of mass, we find that the matching procedure is considerably simplified when we express the Lagrangian in terms of the \textit{intrinsic} contact elements of the inner binary, that is, those defined in the rest frame of the center of mass of the inner binary. This is due to the fact that an appropriate reference frame is needed to disentangle the gravitational field from the multipole moments of an object, as discussed in \cite{Levi:2015msa,Kuntz2021}. For this reason, we will express the terms that appear in the effective Lagrangian, a functional evaluated in the global three-body rest frame $\mathsf{R}$, in terms of the coordinates of the inner binary rest frame $\mathsf{R}'$, so as to obtain a functional depending on the \textit{intrinsic} contact elements of the inner binary.
As studied in \cite{Kuntz2021}, terms in the effective Lagrangian of order up to $ \varepsilon^{3/2} $ are not affected by this difference in reference frame. However, when computing quadrupole order contributions, we will see that it becomes important to account for such a difference.

While deferring the explicit computations in Appendix~\ref{app:CM}, let us just show here the final relation between the absolute coordinates $(\bm y_1, \bm y_2)$ and the relative ones, $\bm Y_\mathrm{CM}, \bm V_\mathrm{CM}$ (center-of-mass position and velocity of the inner binary to 1PN order) and $\bm r', \bm p'$ (relative distance and momentum in the inner binary rest frame to 1PN order):

\begin{align}\label{eq:x1x2_CM'}
\begin{split}
\bm{y}_1 &=  \mathbf{Y}_\mathrm{CM}  + (X_2 + \delta) \mathbf{r}' + X_2 \big( \bm V_\mathrm{CM} \cdot \bm r' \big) \bigg[ \big(X_1-X_2\big) \frac{\bm p'}{\mu} - \frac{\bm V_\mathrm{CM}}{2} \bigg] \; , \\
 \quad \bm{y}_2 &=  \mathbf{Y}_\mathrm{CM}  + (-X_1 + \delta) \mathbf{r}' - X_1 \big( \bm V_\mathrm{CM} \cdot \bm r' \big) \bigg[ \big(X_1-X_2\big) \frac{\bm p'}{\mu} - \frac{\bm V_\mathrm{CM}}{2} \bigg] \; ,
 \end{split}
\end{align}
where $\delta$ is a 1PN quantity defined by
\begin{equation}
    \delta = - \frac{1}{m} \mathbf{V}_\mathrm{CM} \cdot \mathbf{p}' + \nu (X_1 - X_2) \left( \frac{p'^2}{2 \mu} - \frac{G_N m}{2r'} \right).
\end{equation}
In particular, note that this result implies the following relation between $\bm r = \bm y_1 - \bm y_2$ and $\bm r'$:
\begin{equation}
	\bm r = \bm r' - \big( \bm V_\mathrm{CM} \cdot \bm r' \big) \bigg[ \frac{\bm V_\mathrm{CM}}{2} + \big( X_2-X_1 \big) \frac{\bm p'}{\mu} \bigg] \; .
\end{equation}
The final step is just to express $ \bm r' \,,\,\bm p' $ in terms of the contact elements as in Eq. \eqref{eq:ParamContactElements}.
To avoid clutter, in the rest of the article we will suppress the primed label on $\bm r'$, $\bm p'$.


\subsection{Averaging the Lagrangian}

In this Section, we start the computation of the quadrupolar Lagrangian by carrying out Step 2 described in Section~\ref{sec:summary}. Specifically, we expand the Lagrangian~\eqref{eq:L1PN_unexpanded} in multipoles around the center-of-mass of the inner binary, using the formulas given in Eq.~\eqref{eq:x1x2_CM'}. When carrying out computations up to order $v^2 \varepsilon^{5/2}$, it is crucial to include corrections to the leading order averaging procedure besides the quadrupolar order of the multipole expansion. These are due to deviation from perfect adiabaticity, i.e. small changes of slowly evolving quantities over the course of the orbital period, and to backreaction of the quickly oscillating terms on the long time-scale dynamics. In order to simplify the presentation, we leave the detailed analysis of these corrections to App. \ref{app:beyond_adiab}. Here we only remark that such corrections come in the form of the so-called cross terms: in this case either PN corrections to 0PN quadrupole terms or quadrupolar corrections to 1PN terms of the Lagrangian.

To give a separate treatment of these different contributions, we split the final averaged Lagrangian for the inner binary coupled to an external gravitational field, $ \mathcal{L}_{\mathrm{quad},12}^{\leq v^2\varepsilon^{5/2}} $, in two terms: the quadrupole term coming from the multipole expansion, computed to 1PN order, and the cross-terms induced by corrections to the leading order averaging procedure, computed in App. \ref{app:beyond_adiab}:
\begin{align}\label{eq:Lquad0LS}
\mathcal{L}_{\mathrm{quad},12}^{\leq v^2\varepsilon^{5/2}}= \langle\mathcal{L}_{\mathrm{quad}}^{(0)} \rangle+\big\langle\mathcal{L}_S\big\rangle\;.	
\end{align}
From this Lagrangian, integrating out the potential gravitons exchanged with the third body and averaging over the outer orbit, in the next sections we will obtain the long-timescale Lagrangian $ \mathcal{L}_{\mathrm{quad}}^{\leq v^2\varepsilon^{5/2}} $ presented in Step 4 of Section \ref{sec:summary}.

Starting with the quadrupole term of the multipole expansion, to order $ v^2\varepsilon^{5/2} $ included, we obtain the expression
\begin{align} \label{eq:quad_before_average}
\begin{split}
\mathcal{L}_\mathrm{quad}^{(0)} =& - \frac{1}{2} \mu r^i r^j \partial_i \partial_j \tilde \phi \bigg[ 1 - \frac{G_N m}{r} (1-2\nu) + \frac{3}{2} \big( 2(X_2-X_1) \mathbf{V}_\mathrm{CM} \cdot \mathbf{v} + (1-3\nu) v^2 \big) \bigg.\\\bigg.&+ \big( X_1-X_2 \big) \bm V_\mathrm{CM} \cdot \bm r \big( r^i v^j+r^j v^i \big) \bigg] + \frac{1}{2} \mu r^i r^j \partial_i \partial_j \tilde A_k \big[ V_\mathrm{CM}^k + (X_2-X_1) v^k \big] \; ,
\end{split}
\end{align}
where all the relevant quantities used in this equation have been defined in Section~\ref{sec:CM} and $\bm r\,,\,\bm p$ are the relative position and momentum vectors in the instantaneous rest frame of the inner binary center of mass.\footnote{In this expression, we can substitute $ \bm v = \bm p/\mu $.}
We have made several simplifications in order to get to Eq.~\eqref{eq:quad_before_average}. First, we have dropped terms nonlinear in $\tilde \phi$ as well as $V_\mathrm{CM}^2$ corrections as they would be of order $v^2 \varepsilon^3$ according to the power-counting rules of our theory, once the external gravitational field is integrated out. These terms would be relevant when considering the octupolar order in the center-of-mass expansion, which is beyond the scope of this paper. Second, although at this point the coupling $\partial_i \partial_j \tilde A_k  V_\mathrm{CM}^k$ is of $\varepsilon^{5/2} v$ order, it cannot contribute to the final $\varepsilon^{5/2} v^2$ Lagrangian when $\tilde A_k$ is been integrated out. This is because the lowest-order coupling of $\tilde A_k$ to $m_3$ involves also $v_3$, so that it brings another $\varepsilon^{1/2}$ factor once we integrate out $\tilde A_k$, due to the scaling $  v_3\sim \varepsilon^{1/2} v$. Therefore we can ignore the term $\partial_i \partial_j \tilde A_k  V_\mathrm{CM}^k$ in the present analysis. Finally, note that the use of intrinsic relative position vectors, defined in the inner binary instantaneous rest frame as explained in Section~\ref{sec:CM}, manifests itself in the last factor that multiplies $ \partial_i\partial_i\tilde{\phi} $.

Following the method of near-identity transformations, presented in Appendix \ref{app:toymodel}, we can implement the averaging procedure and eliminate the short time-scale dynamics from the Lagrangian. We leave to Appendix \ref{app:beyond_adiab} the corrections due to breaking of adiabaticity and backreaction of short modes, $\langle\mathcal{L}_S\rangle$ in Eq. \eqref{eq:Lquad0LS}, while here we average Eq.~\ref{eq:quad_before_average} over one orbit of the inner binary, using the definition of contact elements given in Eq.~\eqref{eq:DefContactElements}. We obtain:
\begin{align} \label{eq:quadrupole_naive}
\begin{split}
\langle \mathcal{L}_\mathrm{quad}^{(0)} \rangle &= - \frac{\mu a^2}{4} \bigg[ (1+4e^2) \alpha^i \alpha^j + (1-e^2)\beta^i \beta^j \\
&+ \frac{G_N m}{2 a} \big( (1-5\nu + 4e^2(2\nu-1)) \alpha^i \alpha^j +(1-5\nu)(1-e^2) \beta^i \beta^j \big) \bigg]  \partial_i \partial_j \tilde \phi \\
&  - \frac{1}{2}  (X_2-X_1) a e \; \epsilon_{l i k} J^l \alpha^j \big( \partial_i \partial_j \tilde A_k - 4 V_\mathrm{CM}^k \partial_i \partial_j \tilde \phi  \big) \; ,
\end{split}
\end{align}
where we recall that $\bm J$ is the angular momentum of the binary defined in Eq.~\eqref{eq:def_J_Omega}.  

At this point we can compute the whole contribution $ \mathcal{L}_{\mathrm{quad},12}^{\leq v^2\varepsilon^{5/2}} $ by adding the result from our analysis in Appendix \ref{app:beyond_adiab}, i.e. Eq.~\eqref{eq:L1_crossTerms}, to the contribution of Eq. \eqref{eq:quadrupole_naive}:
\begin{align} \label{eq:quadrupole_full}
\begin{split}
\mathcal{L}_{\mathrm{quad},12}^{\leq v^2 \varepsilon^{5/2}} &= - \frac{1}{2} Q_E^{ij}  \partial_i \partial_j \tilde \phi  - \frac{1}{2}  (X_2-X_1) a e \; \epsilon_{l i k} J^l \alpha^j \big( \partial_i \partial_j \tilde A_k - 4 V_\mathrm{CM}^k \partial_i \partial_j \tilde \phi  \big) \; .
\end{split}
\end{align}
Here the traceless "electric-type" quadrupole moment is given to 1PN order by
\begin{equation}\label{eq:Qij1PN}
Q_E^{ij} = \frac{\mu a^2}{2} \bigg( f_\alpha(e) \alpha^i \alpha^j + f_\beta(e) \beta^i \beta^j - \frac{f_\alpha(e) + f_\beta(e)}{3} \delta^{ij}  \bigg) \; ,
\end{equation}
and the two functions of the eccentricity read
\begin{align}
f_\alpha(e) &= 1 + 4e^2 - \frac{G_N m}{2 a\big(1-e^2+\sqrt{1-e^2}\big)} \bigg[17 + 13 \sqrt{1-e^2} +5\nu \big(1+\sqrt{1-e^2}\big) \nonumber \\
&+ e^2 \big(56+15\sqrt{1-e^2} -\nu(13+8 \sqrt{1-e^2}) \big) + 4e^4 \big(3+2\nu \big)\bigg] \; , \\
f_\beta(e) &= 1 -e^2 - \frac{G_N m}{2 a\big(1-e^2+\sqrt{1-e^2}\big)} \bigg[13 + 17 \sqrt{1-e^2} +5\nu \big(1+\sqrt{1-e^2}\big) \nonumber \\
&+ e^2 \big(31+18\sqrt{1-e^2} -5 \nu(2+ \sqrt{1-e^2}) \big) + e^4 \big(5\nu -9 \big)\bigg] \; .
\end{align}
The first term of the Lagrangian~\eqref{eq:quadrupole_full} contains both Newtonian $\varepsilon^2$ and PN $v^2 \varepsilon^2$ scalings, while its second term proportional to the difference of masses contains only the PN $v^2 \varepsilon^{5/2}$ scaling.
Note that, in order to remove the trace from the quadrupole moment, we have made use of the equation of motion $\partial_i \partial^i \tilde \phi = 0$.\footnote{\label{f:EOMs}Even though we will eventually be interested in off-shell potential gravitons exchanged between the inner binary and the third body, using this on-shell condition does not alter the effective action, similarly to what is discussed in \cite{Goldberger:2007hy}, since it amounts to neglecting a contact term.}
Note also that the osculating elements used in these equations are the \textit{intrinsic contact} elements, defined in the rest frame of center of mass of the inner binary, and not Newtonian orbital elements. The explicit difference between these two sets of osculating elements is detailed in Appendix~\ref{app:contact}, see Eq.~\eqref{eq:shift_contact_osculating}. As explained there, this difference is small in the sense that at any time, the contact elements differ from the osculating elements by a 1PN quantity. Despite the small difference, Eq.~\eqref{eq:shift_contact_osculating} implies that while the contact element $a$ is constant by virtue of our averaging procedure, see Appendix~\ref{app:semimajor}, the respective orbital element features post-Newtonian variations. This confirms the findings of~\cite{PhysRevD.89.044043,Will:2014wsa}, who showed for the first time that cross-terms induce variations in the (orbital) semimajor axis $a$. However, our Lagrangian formalism allows us to assert that these variations always stay small over time and cannot accumulate over a long timescale to induce large variations in $a$, while this was left as an open question in previous works~\cite{PhysRevD.89.044043,Will:2014wsa}.

\subsection{Matching} \label{sec:matching}

We will now express the quadrupolar coupling~\eqref{eq:quadrupole_full} in a gauge-invariant way. As shown in e.g.~\cite{goldberger_gravitational_2010, porto_effective_2016}, the quadrupolar part of the effective action can be written in terms of two interactions, the electric-type and magnetic-type quadrupole terms:
\begin{align}
\begin{split} \label{eq:nonminimal_quadru}
\mathcal{S}_\mathrm{quad} =  - \frac{1}{2} \int \mathrm{d} \tau E_{ij} Q_E^{ij} + \frac{2}{3} \int \mathrm{d} \tau B_{ij} Q_B^{ij} \; ,
\end{split}
\end{align}
where $Q_E^{ij}$ and $Q_B^{ij}$ are the electric-type and magnetic-type quadrupole moments of the source, coupled to the corresponding parts of the Weyl tensor $C_{\mu \nu \alpha \beta}$:
\begin{align}
\begin{split}
E_{\mu \nu} &= C_{\mu \alpha \nu \beta} V_\mathrm{CM}^\alpha V_\mathrm{CM}^\beta \; , \\
B_{\mu \nu} &= \frac{1}{2} \epsilon_{\mu \alpha \beta \sigma} C^{\alpha \beta} {}_{\nu \rho} V_\mathrm{CM}^\sigma V_\mathrm{CM}^\rho \; .
\end{split}
\end{align}
Note that in Eq.~\eqref{eq:nonminimal_quadru} these tensors have been projected to a locally flat frame.

Being only interested in terms of order $v^2 \varepsilon^{5/2}$, we can expand $E_{\mu \nu}$ and $B_{\mu \nu}$ to linear order in the gravitational fields $\tilde \phi$ and $\tilde A_i$. Furthermore, at this order we can also ignore the interactions involving  $\dot{\tilde A}_i$ as well as both $V_\mathrm{CM}$ and $\tilde A_i$ together. Finally, we will also make use of the equations of motion for the external fields which read $\partial_i \partial^i \tilde \phi = \partial_i \partial^i \tilde A_j = 0$,\footnote{As mentioned in Footnote \ref{f:EOMs}, even if we will eventually be interested in off-shell gravitons we still can use the equations of motion, which amounts to neglecting a contact term.} and of the gauge condition on $\tilde A^i$ which is $\partial_i \tilde A^i = -4 \dot{ \tilde \phi}$~~\cite{Kol:2007bc, Kol:2010si}. Thus, the tensors are given by
\begin{align}
E_{ij} &= \partial_i \partial_j \tilde \phi \; , \\
B_{ij} &= \frac{1}{2} \epsilon_{m n (i} \partial_{j)} \big[\partial^m \tilde A^n + 4 V_\mathrm{CM}^m \partial^n \tilde \phi \big] \; . \label{eq:Bij_expanded}
\end{align}

The resulting magnetic-type quadrupole reproduces the coupling to $\tilde A$ and $\tilde \phi$ proportional to the difference of masses $X_1-X_2$  that we found previously in Eq.~\eqref{eq:quadrupole_full}~\footnote{In order to get the coupling in Eq.~\eqref{eq:quadrupole_full} from the two equations~\eqref{eq:Bij_expanded} and~\eqref{eq:QBij}, one has to make use of the gauge condition on $\tilde A_i$ and $\tilde \phi$ and the fact that 
\begin{equation*}
\frac{\mathrm{d} \tilde \phi}{\mathrm{d} t} = \dot{\tilde \phi} + V_\mathrm{CM}^i \partial_i \tilde \phi
\end{equation*}
so that this total derivative can be ignored from the vertex.

 }:
\begin{equation}
Q_B^{ij} = \frac{3}{2} (X_1 - X_2) a e J^{(i} \alpha^{j)} \label{eq:QBij} \; ,
\end{equation}
This is consistent with the standard definition of the magnetic-type quadrupole \begin{align}
	Q_B^{ij} = \epsilon^{m n (i} \sum_A m_A x_A^{j)} x_A^m v_A^n \; ,
\end{align} 
see e.g. \cite{porto_effective_2016}, once averaged over one orbit of the inner binary system, thus giving a strong check of the validity of our decomposition. Note that the use of the relative variables defined in the rest-frame of the inner binary, as explained in Section~\ref{sec:CM}, is crucial to obtain the correct magnetic-type interaction. Had we used relative variables naively defined in the three-body rest frame (as it is the case in other studies~\cite{PhysRevD.89.044043,PhysRevLett.120.191101,Will:2014wsa, Lim:2020cvm}), we would not have achieved this separation between a quadrupole moment intrinsic to the inner binary coupled to the external gravitational field. Instead, we would have obtained a quadrupole moment which depends on the center-of-mass velocity of the inner binary. This undesirable feature has been avoided thanks to our choice of variables.

To recap our results, we obtain the following action for the binary system treated as a point-particle up to quadrupole order:
\begin{equation} \label{eq:action_quadrupole_verySimple}
S = \int \mathrm{d} \tilde \tau \bigg[ - \mathcal{E}_\mathrm{1PN} + \frac{1}{2} J_{\mu \nu} \Omega^{\mu \nu} - \frac{1}{2} E_{ij} Q_E^{ij} + \frac{2}{3} B_{ij} Q_B^{ij} \bigg] \; ,
\end{equation}
where $Q_B^{ij}$ and $Q_E^{ij}$ have been defined in Eqs.~\eqref{eq:QBij} and~\eqref{eq:Qij1PN}, $d \tilde \tau = dt\sqrt{-\tilde g_{\mu \nu} V_\mathrm{CM}^\mu V_\mathrm{CM}^\nu}$ and the first two terms represent the Lagrangian of the inner binary to dipole order, with $\mathcal{E}_\mathrm{1PN}$ its energy to 1PN order, $J_{\mu \nu}$ its spin tensor and $\Omega^{\mu \nu}$ its angular velocity rotation vector. 
These quantities have been computed in \cite{Kuntz2021}, see Eq.~(45) there. We can now move on to Step 3 of Section \ref{sec:summary} and integrate out the external fields $\tilde \phi$, $\tilde A_i$ in the presence of a third point-particle $m_3$.

\section{Double-averaged Lagrangian up to order $v^2 \varepsilon^{5/2}$} \label{sec:double_averaged_Lagrangian}

\begin{figure}
	\centering
	\subfloat[]{
		\includegraphics[width=0.2\columnwidth]{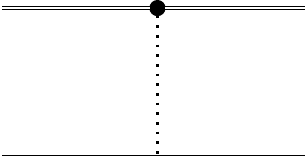}
	}\hspace{1em}
	\subfloat[]{
		\includegraphics[width=0.2\columnwidth]{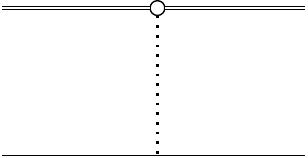}
	}\hspace{1em}
	\subfloat[]{
		\includegraphics[width=0.2\columnwidth]{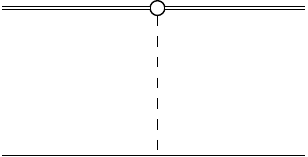}
	}\hspace{1em}
	
\caption{Feynman diagrams obtained by integrating out the outer fields $\tilde \phi$, $\tilde{\bm A}$ and contributing to the quadrupolar Lagrangian~\eqref{eq:Lquad_final_before_avg} up to order $v^2 \varepsilon^{5/2}$. The double line represents the inner binary system, while the single line stands for the third particle $m_3$. The black [white] dot represent the insertion of an "electric-type" ["magnetic-type"] quadrupolar coupling from Eq.~\eqref{eq:action_quadrupole_verySimple}. The dotted line represents the exchange of a scalar $\tilde \phi$, while the dashed line stands for the exchange of a vector $\tilde{\mathbf{A}}$.}
\label{fig:feynDiagr_quad}
\end{figure}

\begin{table}
\center
\begin{tabular}{|c|c|}
\hline
Operator & Rule  \\
\hline
$\displaystyle{ - \frac{1}{2} Q_E^{ij} \partial_i \partial^j \tilde \phi}$ & $J_3^{1/2} v^2 \varepsilon^{2} $  \vphantom{\bigg[} \\
\hline
$\displaystyle{ \frac{1}{3}  Q_B^{ij}   \epsilon_{m n (i} \partial_{j)} \partial^m \tilde A^n}$ & $J_3^{1/2} v \varepsilon^2  $  \vphantom{\bigg[} \\
\hline
$\displaystyle{ \frac{4}{3}  Q_B^{ij}   \epsilon_{m n (i} \partial_{j)} V_\mathrm{CM}^m \partial^n \tilde \phi}$ & $J_3^{1/2} v^2 \varepsilon^{5/2}$  \vphantom{\bigg[} \\
\hline

\end{tabular}
\caption[Power-counting rules]{Power-counting rules for the quadrupolar vertices contained in the effective point-particle action~\eqref{eq:action_quadrupole_verySimple} up to 1PN order, with $J_3 = (G_N M^3 a_3)^{1/2}$,  $v^2 = G_Nm/a$ and $\varepsilon = a/a_3$. The rules are obtained using the scaling presented in Section IV A of \cite{Kuntz2021}. For convenience, the integral over time is not displayed, although it should be included to obtain a dimensionless rule. Furthermore, in the main text we will ignore the $J_3$ factors when discussing the scaling of an operator, since in the end the Lagrangian will always be proportional to $J_3$ (terms not proportional to the angular momentum represent true quantum loops whose contribution to the dynamics is completely negligible in the NRGR formalism~\cite{goldberger_effective_2006})}
\label{table:power_counting}
\end{table}

In the previous sections we have derived the Lagrangian  $ \mathcal{L}_{\mathrm{quad},12}^{\leq v^2\varepsilon^{5/2}} $ describing the dynamics of the inner binary coupled to an external gravitational field over time-scales longer than the inner orbital period, up to order $ v^2\varepsilon^{5/2} $. We now take into account the presence of the third body. This implies that we need to integrate out potential gravitons exchanged between the inner binary and the third body, and in turn adds a new time-scale to the problem: the orbital period of the outer orbit. Being interested in the evolution on much longer time-scales, we will repeat the averaging procedure carried out for the inner binary, starting from quantities that are already averaged with respect to the period of the inner binary.

The action of the third point-particle is given by:
\begin{equation}
\mathcal{L}_3 = - m_3 \sqrt{- \tilde g_{\mu \nu} v_3^\mu v_3^\nu } \; ,
\end{equation}
and we integrate out the gravitational field $\tilde g_{\mu \nu}$. Since the corresponding Lagrangian up to dipole order $v^2 \varepsilon^{3/2}$ has already been described in \cite{Kuntz2021}, we will concentrate on the quadrupolar contributions\footnote{Being interested in the dynamics of contact elements, we do not keep track of the possible coupling of the three body system itself to an external gravitational field.}. From Eq.~\eqref{eq:action_quadrupole_verySimple}, we see that three new vertex appear at quadrupolar order, whose power-counting rules are summarized in Table~\ref{table:power_counting}. Integrating out the external fields $\tilde \phi$, $\tilde{\bm A}$ as shown in the Feynman diagrams of Figure~\ref{fig:feynDiagr_quad}, we find
\begin{equation}\label{eq:Lquad_final_before_avg}
\mathcal{L}_\mathrm{quad}^{\leq v^2 \varepsilon^{5/2}} = \frac{G_N m_3}{2 R^3} \big(3 N^i N^j - \delta^{ij} \big) \bigg[ Q_{E, ij} +  4 (X_2-X_1) a e \epsilon_{l k i} J^l V^k \alpha_j \bigg] \; ,
\end{equation}
where we have moved to the center-of-mass frame of the triple system by setting $\bm V_\mathrm{CM} = X_3 \bm V$, $\bm v_3 = - X_\mathrm{CM} \bm V$, and we recall that $Q_E^{ij}$ has been defined in Eq.~\eqref{eq:Qij1PN}. This Lagrangian contains terms with three different scalings: Newtonian quadrupolar ($\varepsilon^2$) in the Newtonian part of $Q_E^{ij}$, 1PN quadrupolar ($v^2 \varepsilon^2$) in the 1PN part of $Q_E^{ij}$, and magnetic-type quadrupolar ($v^2 \varepsilon^{5/2}$) in the second term proportional to the difference of masses. 

We can now implement the second step of averaging and eliminate the dynamics on time-scales shorter than the period of the outer orbit by means of new near-identity transformations. We thus obtain the doubly averaged Lagrangian:

\begin{mybox}
\begin{align} \label{eq:Lquad_final}
\left\langle\mathcal{L}_\mathrm{quad}^{\leq v^2 \varepsilon^{5/2}}\right\rangle =& \frac{3 G_N m_3}{4 a_3^3 (1-e_3^2)^{3/2}} Q_E^{ij} \big(\alpha_3^i \alpha_3^j + \beta_3^i \beta_3^j \big) \\\nonumber &+ \frac{3 (X_2-X_1)}{X_\mathrm{CM}} \Omega_\mathrm{prec}  \frac{a}{a_3} \frac{e e_3}{1-e_3^2} \bigg( (\bm J \times \bm \alpha) \cdot (\bm J_3 \times \bm \alpha_3) - 2 (\bm J \cdot \bm J_3) (\bm \alpha \cdot \bm \alpha_3)  \bigg)  \; ,
\end{align}
\end{mybox}
where $\Omega_\mathrm{prec}$ is the precession frequency already defined in Eq.~\eqref{eq:omegaPrec}.
This completes our derivation of the double-averaged Lagrangian up to order $v^2 \varepsilon^{5/2}$. 
Note that in doing this last average, we did not need to compute corrections due to deviation from adiabaticity and backreaction of short modes, since these effects, similarly to what discussed in Appendix \ref{app:beyond_adiab}, would contribute only to quadrupole times $ v_3^2 $ order which means starting from octupole order, since $ v_3\sim\e^{1/2}v $.

Starting from this result, the LPE for the long time-scale evolution of both the inner and outer orbits are obtained by taking the relevant derivatives of the Lagrangian as shown in Eqs.~\eqref{eq:dot_a}-\eqref{eq:dot_Omega}. Note that the angular dependence of the quadrupolar $v^2 \varepsilon^2$ terms is quite similar to the Newtonian quadrupole at $\varepsilon^2$, so that we expect that these $v^2 \varepsilon^2$ terms will not induce a qualitatively different behavior in the long-term evolution of the system. On the other hand, the angular structure of the $v^2 \varepsilon^{5/2}$ terms is more involved and somewhat similar to the Newtonian octupole, which means that similarly to the octupole such terms can give rise to new behaviors at long times (see e.g.~\cite{10.1093/mnras/stt302,Naoz_2016} concerning the influence of octupole terms in the Kozai-Lidov problem). We will describe these new behaviors in the next Section, where we will numerically solve the LPE for the long time-scale dynamics obtained from the quadrupolar Lagrangian up to order $v^2 \varepsilon^{5/2}$. 

Before moving on, let us compare our result to the ones already present in the literature. As we stated in the introduction, we are aware of only four works which tackled the task of computing 1PN quadrupolar terms including the effects due to deviations from the adiabatic approximation and to backreaction of quickly oscillating modes. In three of them, a circular outer orbit is assumed~\cite{PhysRevD.89.044043,PhysRevLett.120.191101,Will:2014wsa}. As shown by Eq.~\eqref{eq:Lquad_final}, this causes to neglect the presence of magnetic-type quadrupolar terms of order $v^2 \varepsilon^{5/2}$, which can lead to new interesting behaviors in the long-term evolution of the system as we show in Section~\ref{sec:numerical}. The explicit comparison between the LPE obtained in these works and our result is complicated by the fact that we do not use the exact same averaging procedure, see the comments at the end of Appendix~\ref{app:contact}. On the other hand, the work in~\cite{Lim:2020cvm} reports a puzzling result. Indeed, it describes the effect of a so-called libration cross-term in the equations of motion which, once translated to the Lagrangian point of view, would scale as $v^2 \varepsilon^{1/2}$ within our power-counting rules. Such a term is absent from our derivation and we believe that it should not be present~\footnote{Note that before averaging over the inner orbit, the Lagrangian indeed contains terms scaling as $v^2 \varepsilon^{1/2}$ which are also dependent on the center-of-mass definition. However, one can check that even with the Newtonian definition of the center-of-mass used in~\cite{Lim:2020cvm}, the average of these $v^2 \varepsilon^{1/2}$ terms over the inner orbit still vanishes.}.

\section{Numerical solution to the LPE} \label{sec:numerical}

In this Section, we will numerically integrate the equations of motion stemming from the averaged Lagrangian~\eqref{eq:totLagrangian} for both inner and outer orbits given some initial conditions. A systematic exploration of the (huge) parameter space, as has been done in e.g.~\cite{Naoz_2013} with lower-order perturbations in the Lagrangian, is out of the scope of this paper. Instead, we content ourselves with showing that the quadrupolar terms derived in this paper can have some non-trivial consequences on the long-term evolution of relativistic three-body \hie systems.

Varying the total averaged Lagrangian~\eqref{eq:totLagrangian} over planetary elements as described in Appendix~\ref{app:LPE}, we obtain the so-called Lagrange Planetary Equations (LPE) which dictate the evolution of orbital elements over long timescales. For the inner orbit, they are given by
\begin{align} \label{eq:LPE_mainText}
\dot a &= 0 \; ,\\
\dot e &= - \sqrt{\frac{1-e^2}{G_N m a e^2}} \frac{\partial  \mathcal{R}}{\partial \omega} \; , \\
\dot \iota& = - \frac{1}{\sqrt{G_N m a(1-e^2)} \sin \iota} \frac{\partial  \mathcal{R}}{\partial \Omega}
+ \frac{\cos \iota}{\sqrt{G_N m a(1-e^2)} \sin \iota} \frac{\partial  \mathcal{R}}{\partial \omega} \; , \\
\dot \omega &= \sqrt{\frac{1-e^2}{G_N m a e^2}} \frac{\partial  \mathcal{R}}{\partial e} - \frac{\cos \iota}{ \sqrt{G_N m a(1-e^2)} \sin \iota} \frac{\partial \mathcal{R}}{\partial \iota} \; ,  \\
\dot \Omega &=  \frac{1}{\sqrt{G_N m a(1-e^2)} \sin \iota} \frac{\partial \mathcal{R}}{\partial \iota} \; , 
\end{align}
where $\mathcal{R} = \delta \mathcal{L}/\mu$ is the so-called \textit{perturbing function}, with $\delta \mathcal{L}$ containing all terms beyond the first two kinetic parts in the total Lagrangian~\eqref{eq:totLagrangian}. The LPE for the outer orbit are obtained by replacing all inner quantities with outer ones. However, one has to be careful to replace the mass $m$ with the sum $m_3 + \mathcal{E}$, with $\mathcal{E} = m - G_N m/2a$, as was emphasized below Eq.~\eqref{eq:totLagrangian}.

Until now, we have not specified any orientation for our reference frame centered on the total center-of-mass of the three-body system. A straightforward application of Noether's theorem to the averaged Lagrangian~\eqref{eq:totLagrangian} gives that the total momentum, $\bm J + \bm J_3$, is conserved. We can thus follow the conventional elimination of the nodes procedure, see e.g.~\cite{Naoz_2016}, to choose the orientation of axis with the z axis parallel to $\bm J + \bm J_3$, in which the following relations between planetary elements hold true:
\begin{equation}
\Omega_3 = \Omega + \pi \; , \quad \Vert \bm J \Vert \sin \iota = \Vert \bm J_3 \Vert \sin \iota_3 \label{eq:EliminationOfNodes} \; .
\end{equation}
This allows us to eliminate two variables from the eight dynamical variables that we are solving for, by e.g. expressing all quantites only in terms of $\Omega$ and $\iota_\mathrm{tot} = \iota + \iota_3$. 
As a relevant remark, let us highlight two subtleties concerning the elimination of nodes. First, as discussed in~\cite{10.1093/mnras/stt302}, one can not directly use the relations~\eqref{eq:EliminationOfNodes} at the level of the Lagrangian or the Hamiltonian, because this would mean using a consequence of the equations of motion (conservation of angular momentum) in the Lagrangian itself, which can lead to wrong results. Indeed, many studies (including the original one of Kozai~\cite{1962AJ.....67..591K}) concluded that the z-components of both angular momentum are conserved \textit{independently}, which is incorrect as shown in~\cite{10.1093/mnras/stt302}. Second, note that the conserved angular momentum is defined with the contact elements, which differ from the osculating elements at 1PN order as we highlight in Appendix~\ref{app:contact}. In other words, the angular momentum defined with osculating elements does feature variations at 1PN order, while the one defined with contact elements is a constant. Thus, one cannot eliminate the nodes by following the conventional procedure if one uses the osculating elements instead of the contact ones.\footnote{The fact that $\dot \Omega \neq \dot \Omega_3$ for osculating elements was already noticed in Ref.~\cite{Lim:2020cvm}. In comparison, our discussion adds that the elimination of nodes can be carried out consistently at the level of contact elements.}
As an independent check of this procedure, our numerical simulation confirms that the projection of the angular momentum on the z-axis, $\Vert \bm J \Vert \cos \iota + \Vert \bm J_3 \Vert \cos \iota_3$, is conserved through the evolution.


\begin{figure}
\includegraphics[width=0.9\columnwidth]{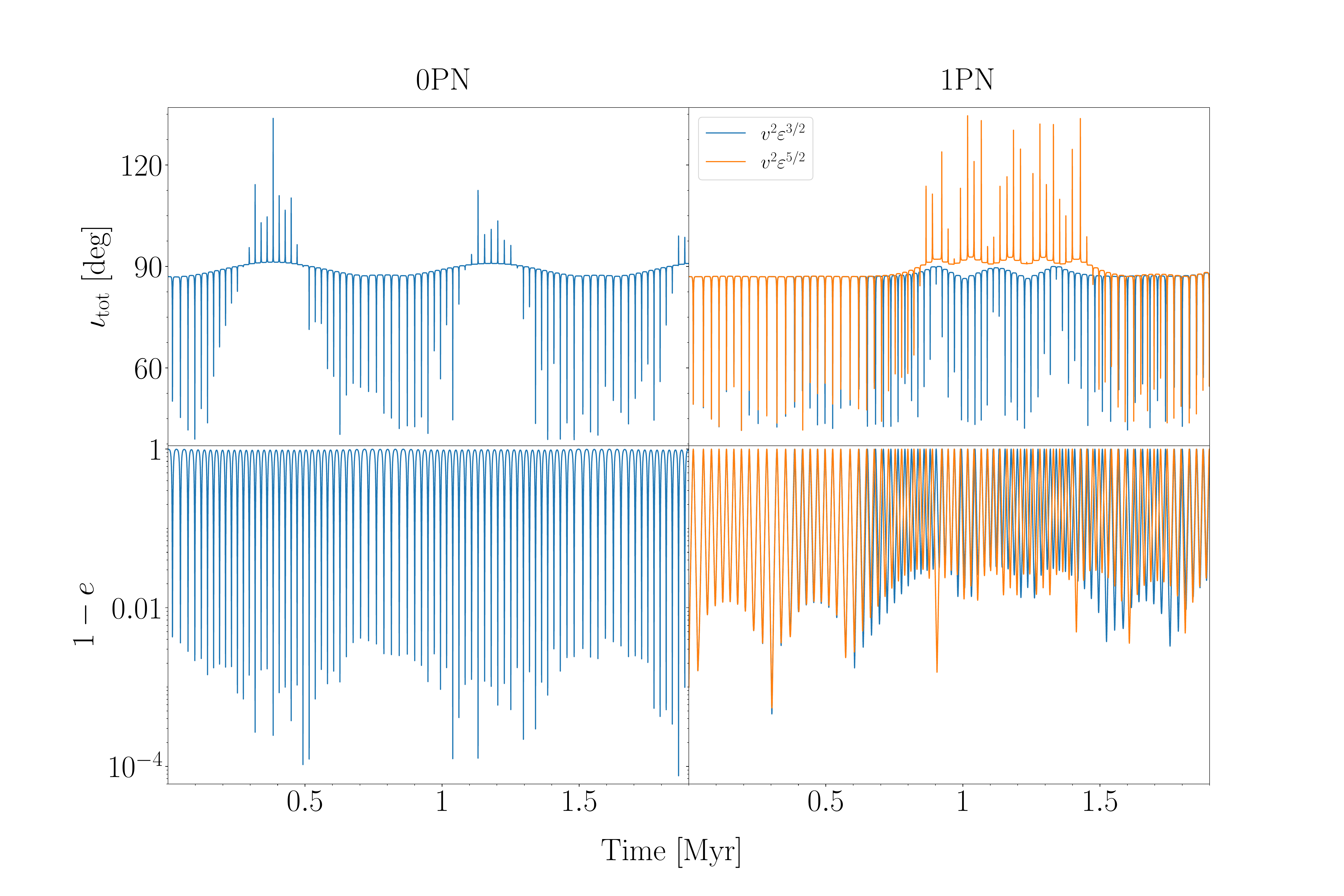}
\caption{\textit{Impact of the quadrupole-1PN terms on the evolution of a three-body system}. We solve the LPE~\eqref{eq:LPE_mainText} with the following parameters for a three-body system: $m=50 M_\odot$, $\nu = 0.15$, $a=5$AU, $m_3=50 M_\odot$, $a_3=350$AU. The initial conditions are $e=0.001$, $e_3=0.7$, $\iota_\mathrm{tot}=87 \degree$, $\omega=240 \degree$, $\omega_3=0 \degree$, $\Omega=0 \degree$. In the left plots we show the evolution of the total inclination $\iota_\mathrm{tot} = \iota + \iota_3$ and the eccentricity $e$ without taking into account PN perturbations (i.e., setting $v^2 =0$ in our power-counting rules), but including Newtonian quadrupolar and octupolar terms as described in e.g.~\cite{Naoz_2016}. The system features orbital flips and eccentricities as high as $1-e \sim 10^{-4}$, which are typical of the octupolar Kozai-Lidov mechanism~\cite{Seto:2013wwa,Stephan_2016,10.1093/mnras/stt302,Naoz_2016}. In the right panels, we show the evolution of the same quantities taking into account higher terms in the PN expansion on top of the Newtonian quadrupolar and octupolar terms. Namely, we include PN terms up to $v^2 \varepsilon^{3/2}$ order (in blue) and up to $v^2 \varepsilon^{5/2}$ order (in orange). The lowest-order PN perturbations (which are well-known and have been studied in the context of three-body dynamics e.g. in~\cite{Ford_2000, Fabrycky_2007, Naoz_2013, PhysRevD.102.104002, Fang:2019mui, 2019ApJ...883L...7L}) generically quench the orbital flips and eccentricities. However, the new $v^2 \varepsilon^{5/2}$ terms computed in this article re-trigger the orbital flips.}
\label{fig:crossTerms}
\end{figure}

We now discuss our numerical solution for a particular set of parameters. We choose the inner binary to be composed of two black holes with total mass $m=20 M_\odot$ and mass ratio $\nu = 0.15$, while the outer perturber has a mass $m_3=50 M_\odot$. The inner semimajor axis is $a= 5$AU, and the outer one is $a_3 = 350$AU. Such values are typical for black holes in dense nuclear clusters~\cite{Randall:2018nud, Leigh_2017, Stephan_2016,Antonini_2016_2}. The initial conditions of the system are described in Figure~\ref{fig:crossTerms}, where we plot the total inclination $\iota_\mathrm{tot} = \iota + \iota_3$ and the eccentricity $e$ as functions of time. Without PN perturbations, the system undergoes flips of inclination and extreme eccentricities due to the octupole effects beyond the Kozai-Lidov mechanism. The presence of lower-order PN terms (up to $v^2 \varepsilon^{3/2}$ i.e. dipole order) quenches the maximal eccentricity as well as the orbital flips. This behavior is well-known in the literature~~\cite{Ford_2000, Fabrycky_2007, Naoz_2013}. However, we find that adding the new 1PN quadrupolar terms that we computed in this article can re-trigger the flips in inclination. This is shown in Figure \ref{fig:crossTerms}. Whether or not this ultimately influences the mechanisms leading to binary mergers in nuclear clusters is left as an interesting question for future work.

As a final comment, note that we have not studied the linear stability of the system, which would require diagonalising the Hessian matrix of the inner and outer contact elements. Even without knowing the eigenvalues of the Hessian matrix, we can appreciate how no unstable direction was hit in our simulation, since amplitude and phases of the oscillations do not grow exponentially. Moreover, we take the persistence of quasi-periodic oscillations as a proxy for the absence of any resonant behaviour (see discussion in Appendix \ref{app:semimajor}). Therefore, the accuracy of the simulation reported is only limited by the growth of higher order terms that we have neglected. These will possibly become of order one after a time which is parametrically larger than the interval explored in the simulation.

\section{Conclusions and Outlook}

In this work we have considered the effects of a distant third body orbiting a tightly bound binary system, forming a hierarchical triple. 
We have derived an effective action that includes relativistic effects up to 1PN order as well as multipole effects up to quadrupole order.

Our EFT approach allows to make use of symmetries to constrain the form of the interactions and makes clear contact with the underlying field theoretic description of gravity.
In particular, starting from an action of three worldlines minimally coupled to gravity, we integrated out the off-shell gravitons that give rise to the binding potential and matched the theory to a system of interacting composite particles. We did this by first integrating potential gravitons binding the inner binary and matching the action of the inner binary to that of a composite particle coupled to an external gravitational field. Then, integrating out potential gravitons with wavelengths comparable to the size of the outer orbit, we obtained an action describing two interacting particles, the inner binary and the third body.
This method allows to build towards a systematic understanding of the long time-scale dynamics away from resonances. 

In practice, deriving the effective action presents a few challenges. For instance, it is important to understand which are the most suitable variables to describe the long time-scale dynamics and the relativistic corrections. In this study we have found that the contact elements defined in the center of mass frame of the inner binary allow to encode in a compact way various PN corrections. Moreover their slowly evolving parts are the quantities that carry only the relevant information to describe the system on long time-scales. Using near-identity transformations, we were able to unambiguously integrate out the effect of fast orbital modes on the dynamics, outlining a procedure that can be used at any order in perturbation theory.
With this approach, we managed to gain insight on the interplay between multipole expansion and relativistic effects. Our main result is that quadrupole-PN cross terms can retrigger orbital flips in spite of PN effects that appear at lower order in the multipole expansion.

We envision a few possible directions for future work.
An obvious possibility is to derive the EFT for the long time-scale dynamics to higher order in either PN or multipole expansion. Several works have derived the three-body potential for gravitational systems in the PM expansion, see e.g. \cite{Loebbert2021}, computing also the first orders of the PN expansion. These results could be used as a starting point to derive the effective action for the long time-scale dynamics of the hierarchical three body problem to higher PN orders. It would also be very interesting to use our effective two-body approach to study gravitational-wave generation in three-body systems; indeed, it is known that a blind application of the quadrupole formula in the center-of-mass frame of a \hie three-body system can lead to unphysical results~\cite{Bonetti_2017}. In contrast, our approach could be used to consistently derive the radiation-reaction force in these systems. 

Another possible direction of work would be to develop a systematic study of the presence or absence of resonances, for given initial data. In this regard, our procedure of integrating out microscopic degrees of freedom in two different steps, despite its simplicity, has the drawback of forcing us to ignore possible resonant behaviours. In contrast, one might integrate out gravitons and perform the averaging in a single step, without inputting from the start the hierarchy between the semimajor axes. For instance, the method adopted in \cite{Jones2022} of performing the hierarchical expansion by using the method of regions might be employed also in the averaging procedure.
In any case, having a result that does not rely on the hierarchy between the two orbits might make possible to detect resonant behaviours as well as to eventually derive our convenient description away from resonances.
Further, it would be interesting to make more explicit the connection between the averaging procedure and the underlying field theory. One possibility could be describing the inner and outer orbit through coherent states of the Kepler problem, see e.g. \cite{Gerry1986}. This might make possible to derive our results diagrammatically, without writing the equations of motion for fast and slow modes.

From the phenomenological point of view, it would be interesting to quantify how these effects alter existing priors on the population of binaries that will be observed through gravitational wave experiments. Even more, our work could be used to better understand which three-body configurations can lead to detectable modifications of gravitational waveforms due to the long time-scale dynamics, and to provide a waveform template incorporating three-body effects. For instance, it has recently been proposed that Kozai-Lidov oscillations may be observable in waveforms~\cite{randallObservingEccentricityOscillations2019, Chandramouli:2021kts}; taking into account the relativistic three-body effects which we computed in this work may be crucial for the parameter estimation of these kind of systems.  

\section*{Acknowledgments}
This research has been partly supported by the Italian MIUR under contract 2017FMJFMW (PRIN2017).

\newpage
\appendix
\section{Averaging through near-identity transformations}\label{app:toymodel}
We here present the averaging procedure that we adopt in our computations, the so-called averaging by near-identity transformations \cite{Murdock}. We do so in the context of a toy model that closely resembles the three-body problem, an angle-periodic system in its canonical form. While in the three body problem there are two angle variables (the mean anomalies of inner and outer orbits) and several slowly evolving variables, we consider a system with one angle variable, $u$, and one slowly evolving variable $x$:
\begin{align}\label{toymodel}
\begin{split}
	\dfrac{d}{dt}x&=\e g_1(x,u)+\e^2g_2(x,u)+\dots \; ,\\
	\dfrac{d}{dt}u&=H(x)+\e h_1(x,u)+\e^2h_2(x,u)+\dots\;,
\end{split}
\end{align}
where $ \e\ll1 $, the dots indicate terms with $ i>2 $, and the functions $ g_i $ , $ h_i $ are periodic in $ {u}\,$ with period $ 2\pi $. Physically, we can think of $ \e $ as a ratio of time-scales, since the slow variable $ x $ has excursions of order $ 1 $ over times that are $ 1/\e $ longer than those over which $ u $ changes its value by an order $ 1 $ factor.
We implement the splitting between fast and slow dynamics using a change of variables:
\begin{align}\label{eq:toymodelansatz}
	\begin{split}
		u(t)&=u_L(t)+\e u_S(x_L,u_L) \; ,\\
		x(t)&=x_L(t)+\e x_S(x_L,u_L)\; ,
	\end{split}
\end{align}
where $ u_S,\,x_S $ encode the short time-scale dynamics and are chosen to be periodic in $u_L$ with period $2\pi$.
Eq. \eqref{eq:toymodelansatz} is called a near-identity transformation, since for $ \e\to 0 $ it reduces to an identity. Physically, this ansatz encodes the fact that fast oscillations will have suppressed amplitude. The strategy that we adopt is to fix the functions $ u_S\,,\, x_S $ in such a way to cancel quickly oscillating terms in the equations \eqref{toymodel} order by order in $ \e $, up to a desired accuracy. Then we truncate the equations, neglecting the higher order corrections that still contain oscillating terms. This will leave a system of equations for $ u_L\,,\,x_L $ that only depends on the long time-scale, up to corrections that are of arbitrarily high order (which are neglected after the truncation).
Plugging the transformation in Eq. \eqref{toymodel} we have, to second order:
\begin{align}\label{eq:nearid}
	\nonumber
	\dot{x}_L+\e(\dot{x}_L\partial_{x_L}+\dot{u}_L\partial_{u_L}){x_S}=& 
		\e g_1(x_L,u_L)+\e^2(x_S\partial_{x_L}+u_S\partial_{u_L})g_1(x_L,u_L)+\e^2g_2(x_L,u_L)+\mathcal{O}(\e^3) \; ,
	\\
	\dot{u}_L+\e(\dot{x}_L\partial_{x_L}+\dot{u}_L\partial_{u_L}){u_S}=&H(x_L)+\e H'(x_L)x_S+\dfrac{\e^2}{2}H''(x_L)x_S^2\\\nonumber+&\e h_1(x_L,u_L)+\e^2(x_S\partial_{x_L}+u_S\partial_{u_L})h_1(x_L,u_L)+\e^2h_2(x_L,u_L)+\mathcal{O}(\e^3)\;,
\end{align}
Now, order by order, we can determine $ x_S\,,\,u_S $ in such a way that the derivatives $\dot{x}_L\,,\,\dot{u}_L$ only depend on $ x_L $, up to terms of order $ \e^3 $:
\begin{align}\label{eq:Uindep}
\begin{split}
	\dot{x}_L=& \e G_1(x_L)+\e^2G_2(x_L)+\mathcal{O}(\e^3) \; ,\\
	\dot{u}_L=&H(x_L)+\e H_1(x_L)+\e^2H_2(x_L)+\mathcal{O}(\e^3)\;.
\end{split}	
\end{align}
In order to obtain this, we write $ x_S =x_S^{(0)}+\e x_S^{(1)}+\dots$ and $ u_S=u_S^{(0)}+\e u_S^{(1)}+\dots $. Then, comparing Eq. \eqref{eq:Uindep} with Eq. \eqref{eq:nearid} we find that at first order it must be:
\begin{align}\label{eq:fasteom}
\begin{split}
	\partial_{u_L}x_S^{(0)}=& \dfrac{1}{H(x_L)}\Big(g_1(x_L,u_L)-G_1(x_L)\Big) \; ,\\
		\partial_{u_L}u_S^{(0)}=& \dfrac{1}{H(x_L)}\Big(h_1(x_L,u_L)+H'(x_L)x_S^{(0)}(x_L,u_L)-H_1(x_L)\Big)\;.
\end{split}	
\end{align}
Integrating the first equation at fixed $ x_L $, we find:
\begin{align}
	x_S^{(0)}(x_L,u_L)=\int_{0}^{u_L}\dfrac{ds}{H(x_L)}\Big(g_1(x_L,s)-G_1(x_L)\Big)+C_0(x_L)\;.
\end{align}
From this, since by assumption $ x_S^{(0)} $ is periodic in $ u_L $ with period $ 2\pi $, we see that it must be:
\begin{align}
	\int_{0}^{2\pi}\dfrac{ds}{H(x_L)}\Big(g_1(x_L,s)-G_1(x_L)\Big)=0\;,
\end{align}
which in turn implies that $ G_1 $ must be chosen to be the average of $ g_1 $:
\begin{align}
	G_1=\langle g_1\rangle_{u_L}=\int_{0}^{2\pi}\dfrac{ds}{2\pi}g_1(x_L,s)\,.
\end{align}
We can further set $ C_0(x_L) $ in such a way that $ x_S^{(0)} $ has zero average. Schematically we will write:
\begin{align}\label{eq:psizero}
	x_S^{(0)}(x_L,u_L)=\dfrac{1}{H(x_L)}\mathrm{Af}_{u_L}\Big(\int^{u_L}ds\mathrm{Af}_s(g_1(x_L,s))\Big),
\end{align}
where the symbol $ \mathrm{Af}_x $ indicates taking the average-free part of the argument with respect to the variable $ x $.

Note that in this approach, the average is defined without ambiguities through an integral over $ u_L $ with $ x_L $ fixed, as a by-product of requiring $ x_S $ to be periodic. This procedure can be thought of as an average over the short time-scale characterizing the evolution of $ u_L $, while keeping fixed the variable $ x_L $. Therefore, we are not truly performing an average over time, but a procedure that is very similar and which ensures all the same an arbitrarily precise approximation. 

Turning to the equation for $ u_S^{(0)} $, we find in the same way that $ H_1(x_L)=\langle h_1\rangle_{u_L} $, since the quantity $ H'x_S^{(0)} $ is average-free (thanks to our choice of $ C_0 $). 
This means that if $ h_1 $ is average-free, then $ \dot{u}_L $ will receive corrections starting at order $ \e^2 $.
We will also choose the constant of integration for $ u_S^{(0)} $ in such a way to make it average-free. Thus it will be:
\begin{align}\label{eq:zetazero}
	u_S^{(0)}(x_L,u_L)=\dfrac{1}{H(x_L)}\mathrm{Af}_{u_L}\Big(\int^{u_L}ds\left[\mathrm{Af}_s(h_1(x_L,s))+H'(x_L)x_S^{(0)}(x_L,s)\right]\Big)\,.
\end{align}
This first order truncation of the approximation, when applied to the newtonian hierarchical three-body problem, gives the Kozai-Lidov long time-scale dynamics.
Turning to second order, we have:
\begin{align}
		\partial_{u_L}x_S^{(1)}=& \dfrac{1}{H(x_L)}\Big(g_2+(x_S^{(0)}\partial_{x_L}+u_S^{(0)}\partial_{u_L})g_1-H_1\partial_{u_L}x_S^{(0)}-G_2(x_L)\Big) \; ,\\\nonumber
		\partial_{u_L}u_S^{(1)}=& \dfrac{1}{H(x_L)}\Big(h_2+H'x_S^{(1)}+\dfrac{H''}{2}{(x_S^{(0)})}^2+(x_S^{(0)}\partial_{x_L}+u_S^{(0)}\partial_{u_L})h_1-H_1\partial_{u_L}u_S^{(0)}-H_2(x_L)\Big)\;,
\end{align}
where $ g_{1,2}\;,\;h_{1,2} $ are evaluated in $ (x_L,u_L) $.
Again, requiring the near-identity transformation to be periodic, we find that $ G_2 $ and $ H_2 $ will be averages of the other terms in the right hand side. Consequently, $ x_S^{(1)} $ and $ u_S^{(1)} $ will be integrals of average-free expressions, and we will be able to fix the constants of integration so as to make $ x_S^{(1)}\,,\,u_S^{(1)} $ average-free as well.
In particular, we find:
\begin{align}\label{eq:sources2}
	G_2=& \int_0^{2\pi}\dfrac{ds}{2\pi}\Big(g_2(x_L,s)+(x_S^{(0)}\partial_{x_L}+u_S^{(0)}\partial_{u_L})g_1(x_L,s)\Big) \; ,\\\nonumber
	H_2=& \int_0^{2\pi}\dfrac{ds}{2\pi}\Big(h_2(x_L,s)+(x_S^{(0)}\partial_{x_L}+u_S^{(0)}\partial_{u_L})h_1(x_L,s)+\dfrac{H''}{2}\big(x_S^{(0)}\big)^2\Big)\;,
\end{align}
where we have dropped the terms $ H_1\partial_{u_L}x_S^{(0)}\,,\, H_1\partial_{u_L}u_S^{(0)}$ and $ H'x_S^{(1)} $, which are average free.
These functions can then be plugged in the equations for $ x_L $ and $ u_L $, and the slow dynamics can be determined up to terms of order $ \e^3 $. Similarly, one can derive the averaging procedure to any order in $ \epsilon $ simply by fixing higher orders of the near-identity transformation, through the functions $ x_S $ and $ u_S $.

\bigskip

The toy model just discussed describes well the procedure we would adopt if we expressed our quantities in terms of the mean anomaly. However, we can only give a closed form expression of the perturbing function in terms of the eccentric anomaly. If we wished to use the variable $ u $ as a proxy for the eccentric anomaly, then we would have to allow a dependence on $ u $ of the function $ H $ driving the leading order evolution of $ u $. It would be then much less straightforward to understand how to define the near identity transformation, since we would have to decide whether we want to retain a $ u_L $ dependence in the functions $ H_i $ driving the evolution of $ u_L $. Moreover, the equations for $ u_S $ would require a more cumbersome integration.

For this reason, we find useful to exploit the averaging procedure derived above, in terms of the mean anomaly, and to simply change integration variables in a consistent way to the eccentric anomaly. If we call $ u $ the mean anomaly and $ \eta $ the eccentric anomaly, then we know that Kepler's equation holds at all times:
\begin{align}\label{eq:kepler}
	u=\eta-e\sin\eta\;,
\end{align}
where $ e $ is the eccentricity, which we can regard as a component of what in general will be the $ x $ vector. This relation ensures that any function periodic in $ u $ is also be periodic in $ \eta $.
Given Kepler's equation, we can perform the near identity transformation and find a relation between $ u_L $ and the eccentric anomaly. Suppose the eccentricity is transformed as:
\begin{align}
	e = e_L(x_L)+\e e_S(x_L,u_L)\;,
\end{align}
then we see that it must hold:
\begin{align}\label{eq:deformedkepler}
	u_L+\e u_S(x_L,u_L)=\eta-(e_L(x_L)+\e e_S(x_L,u_L))\sin\eta\;.
\end{align}
Although the relation between $\eta$ and $u_L$ is very involved, we can avoid complications by performing a near-identity transformation on $ \eta  $ as well:
\begin{align}
	\eta=\eta_L+\e\eta_S(x_L,u_L)\;.
\end{align}
Then, it is possible to fix $ \eta_S $ so as to retain, to all orders, the relation:
\begin{align} \label{eq:def_etaL}
	u_L=\eta_L-e_L\sin \eta_L\;.
\end{align}
This choice makes possible to express the integrands evaluated in $ u_L $ as simple functions of $ \eta_L $. Moreover, it makes clear that at all orders the change of variables will be given by:
\begin{align}
	du_L=d\eta_L (1-e_L\cos \eta_L)\;.
\end{align}
Note that crucially, as a result of integrating at fixed $ x_L $, the Jacobian does not have the denominator factor $ \dot{u}+\dot{e}\sin\eta $ that would appear in $ dt/d\eta $. This is to say that the averaging procedure that we have presented provides a coarse graining of the system in the time coordinate while removing the need of actually performing an integral over time.
\\\\
Before concluding, let us mention other methods to eliminate fast variables besides the method of near-identity transformations. One alternative is the so-called multiple scale analysis, for instance discussed in \cite{Bender,VanKampen}. The idea behind this method is to introduce fictitious variables corresponding to long time-scales, and to determine the dependence on these long time-scales by imposing the cancellation of terms that display a secular growth, i.e. terms growing linearly with time which would break the perturbative expansion early on. As far as we know, the method of near identity transformations is to be preferred over the multiple scale analysis if one is interest in estimating the range of validity in time of the approximate solution \cite{Murdock}. Besides the multiple scale analysis, we quote the method of Von Zeipel transformations, \cite{Hagihara,Naoz_2013}, which works at the level of the Hamiltonian implementing canonical transformations that eliminate the dependence on short time-scale modes, very similarly to what we have done above. Finally, we quote the method of dynamical renormalization group \cite{Galley2017}, which operates in a way similar to the multiple scale analysis, removing secularly growing terms by means of counterterms.

\section{Conservation of contact semi-major axis} \label{app:semimajor}
We now turn to the question of whether the semi-major axes of the two orbits remain constant over long time-scales. Despite the simplicity of this question, to our understanding the answer is quite involved.
There is an intuitive reasoning to argue that the semi-major axes are constant over long time-scales. That is, after the averaging procedure is carried out, the Lagrangian becomes independent on the mean anomaly of the corresponding orbit, therefore making the corresponding conjugate momentum constant. The latter, as shown in Eq. \eqref{eq:def_L_G_H}, depends only on the semimajor axis.
However, when considering PN corrections this argument can only hold for the contact element $ a $, rather than for the orbital element $ \tilde{a} $, see discussion in Appendix \ref{app:contact}. Moreover, as we discuss in Appendix \ref{app:toymodel}, the dynamical variables left after the averaging procedure are the long time-scale modes of the original, full contact elements. Therefore the statement will not hold for the full contact element, but only for its slowly evolving part.

The way to make this intuition rigorous is to implement a canonical transformation of the Hamiltonian, eliminating order by order the dependence on the mean anomalies. This is achieved through the Von Zeipel transformations, as discussed in \cite{Naoz_2013}. The result is that indeed, as long as perturbation theory goes, the semimajor axes remain constant over long time-scales.
\\\\
The systematic control over each order in perturbation theory given by the near identity transformations allows for an independent check of this statement.
Explicitly, using the formulas derived for the toy model \eqref{toymodel}, we can check that, for instance, to second order we expect the semimajor axis to be constant on long timescales. To argue this, we can think of the functions appearing as sources on the right hand side of Eq. \eqref{toymodel} as partial derivatives of the Hamiltonian of the system $ \mathcal{H}=\mathcal{H}_0+\mathcal{H}_1+\dots $ :
\begin{align}
	g_1=-\dfrac{\partial\mathcal{H}_1}{\partial u}\;&,\;g_2=-\dfrac{\partial\mathcal{H}_2}{\partial u} \; ,\\\nonumber H=\dfrac{\partial\mathcal{H}_0}{\partial x}\;&,\;h_1=\dfrac{\partial\mathcal{H}_1}{\partial x}\;,
\end{align}
where we assume $ \mathcal{H} $ to be periodic in $ u $, meaning that the functions $ g_i $ will be average free. This allow us to inspect order by order whether the time derivative of the conjugate momentum to the mean anomaly vanishes or not, by computing the long time-scale source terms $ G_i $, in Eq. \eqref{eq:Uindep}. For instance, it is evident that $ G_1=0 $, due to $ g_1 $ being average free. For $ G_2 $, given in Eq.\eqref{eq:sources2}, determining the answer is less straightforward. Schematically, we have:
\begin{align}
	G_2&=\int_{0}^{2\pi}\dfrac{ds}{2\pi}\left\{\frac{1}{H}\mathrm{Af}\big(\mathcal{H}_1\big)\partial_x\partial_u\mathcal{H}_1-\partial_x\left(\frac{1}{H}\mathrm{Af}\bigg(\int^u\mathrm{Af}(\mathcal{H}_1)\bigg)\right)\partial_u^2\mathcal{H}_1\right\}\;\\\nonumber
	&=\int_{0}^{2\pi}\dfrac{ds}{2\pi}\left\{\partial_x\left(\frac{1}{H}\mathrm{Af}(\mathcal{H}_1)\partial_u\mathcal{H}_1\right)-\partial_u\left[\partial_x\left(\frac{1}{H}\mathrm{Af}\bigg(\int^u\mathrm{Af}(\mathcal{H}_1)\bigg)\right)\partial_u\mathcal{H}_1\right]\right\}\;.
\end{align}
Here we have used that $ H=H(x) $ and that the integrals over $ u $ are performed at fixed $ x $, as dictated by the method of near identity transformations.
In this expression, the total $u$ derivative gives a vanishing contribution thanks to periodicity of the functions, while in the first term we can recognize a total derivative plus an average free term:
\begin{align}
	\frac{1}{H}\mathrm{Af}(\mathcal{H}_1)\partial_u\mathcal{H}_1=\frac{1}{2H}\partial_u(\mathcal{H}_1^2)-\frac{1}{H}\langle\mathcal{H}_1\rangle\partial_u\mathcal{H}_1\;.
\end{align}
These terms give a vanishing contribution to the average, therefore we find
\begin{align}
	G_2=0\;.
\end{align}
Moving to higher orders, we will have to handle increasingly complex expressions. In the end, also from the point of view of near identity transformations, the simplest route might be showing that the long time-scale part of $ x $ and $ u $ are still conjugate variables described by an Hamiltonian. 
\\\\
Despite these results, the perturbative expansion can fail due to resonances between modes of the two orbits. In practice, if the orbits have commensurable periods, then some terms of the expansion can be enhanced by inverse powers of the expansion parameters, usually called small divisors \cite{Murdock}. As already remarked, in the analysis of this work we have discarded such cases by performing the averages over the two orbits independently. Generally however, small divisors will appear at high enough orders in perturbation theory. Their presence will determine a loss of validity of the predictions that we have obtained and a corresponding non trivial evolution of the semimajor axes on time-scales that are parametrically larger than those characterizing the effects described by lower orders in perturbation theory. Instead if the system is studied close to a resonance, then the standard perturbation theory will stop working already from low orders and resonant behaviour will appear early on in the evolution of the system. As an example, the effects of a resonance on the evolution of a hierarchical triple were studied in~\cite{Kuntz:2021hhm}.

\section{Backreaction and deviations form adiabaticity} \label{app:beyond_adiab}

As outlined in Appendix \ref{app:toymodel}, the averaging procedure can be conveniently defined as an average over the values of the slowly evolving part of the mean anomaly, then expressed in terms of the eccentric anomaly as in Eq.~\eqref{eq:def_etaL}.
Here we apply this procedure to the three body problem. 

\subsection{Long-timescale and short-timescale Lagrangians} \label{sec:LS}

In order to describe corrections to the adiabatic approximation, we should return to the Lagrangian~\eqref{eq:L1PN_unexpanded}.  It will be useful to split it between a Newtonian term and a perturbation:
\begin{equation} \label{eq:def_R}
	\mathcal{L} =  \frac{1}{2} v^2 + \frac{G_N m}{r} + \mathcal{R} = \dfrac{(G_Nm)^2}{2 L^2}+ L \dot u + G \dot \omega + H \dot \Omega  + \mathcal{R} \; ,
\end{equation} 
where the perturbing function $\mathcal{R}$ contains the kinetic term of the center of mass as well as any other term beyond the Newtonian two-body interaction of the inner binary. For simplicity we have divided the Lagrangian by $ \mu $, a notation that we will use throughout the appendices. 
The second equality expresses the Newtonian part as a first-order Lagrangian depending on the (osculating) contact elements of the orbit, which are defined in Section~\ref{sec:contact}. It can be checked that this Lagrangian indeed gives the Lagrange Planetary Equations presented in Section~\ref{app:LPE} and which are usually presented in the Hamiltonian formalism, see also~\cite{relativistic_celestial_mechanics, Kuntz2021}.

Finally, the conjugate momenta are given by
\begin{equation} \label{eq:def_L_G_H}
	L = \sqrt{G_N ma}  \; , \; G = L \sqrt{1-e^2} \; , \; H = G \cos \iota \; .
\end{equation} 

Given Eq. \eqref{eq:def_R}, we can perform near-identity transformations for all the contact elements as outlined in Appendix \ref{app:toymodel}. Once we determine the fast oscillating terms in each of the near-identity transformations to a desired order in both $\e $, the ratio of the semimajor axes, and $ v $, the velocity of the bodies, we can simply plug back in the Lagrangian these values, obtaining a classical effective Lagrangian. Once we expand the resulting equations of motion in $ \e $ and $ v $, we will recover the slow dynamics, as in Eq. \eqref{eq:Uindep}. To simplify even more the Lagrangian, we can take its average over the mean anomaly (keeping fixed slow variables, as prescribed by the near-identity transformation). Doing so will not alter the equations of motion for the slowly evolving variables, since the average commutes with variations of the Lagrangian with respect to these variables. It will simply remove the average-free part of the effective Lagrangian, which does not carry dynamical information. As remarked in the previous Appendix, following this procedure at leading order will lead to the long time-scale Kozai-Lidov dynamics. 

More explicitly, considering first the evolution of $ L $ and $ u $ alone, we employ the following two near-identity transformations:
\begin{align}
	\begin{split}
	u=&u_L+u_S(L_L,u_L) \; ,\\
	L=&L_L+L_S(L_L,u_L)\,,
	\end{split}
\end{align}
where the subscripts $ \hphantom{}_L\,,\,\hphantom{}_S $ stand for the long and short time-scale variable respectively, and the $ \hphantom{}_S $ variables are suppressed by powers of both $ \varepsilon $ and $ v $. Using the LPE for $ u $ and $ L $, see Appendix \ref{app:LPE}, we determine the equations for the short time-scale variables as discussed in Appendix \ref{app:toymodel}, finding at lowest order:
\begin{align}
	\partial_{u_L}{L}_S = \dfrac{L_L^3}{(G_Nm)^2}\mathrm{Af}_{u}\Big(\dfrac{\partial\mathcal{R}}{\partial u}\Big) \; , \quad 	\partial_{u_L}{u}_S =  - \dfrac{L_L^3}{(G_Nm)^2}\Big[\mathrm{Af}_{u}\Big(\dfrac{\partial\mathcal{R}}{\partial L}\Big) +\dfrac{3(G_Nm)^2L_S}{L_L^4}\Big]\;,
\end{align}
where, following the near-identity transformation procedure, at leading order we use $ \dot X \simeq \frac{(G_Nm)^2}{L_L^3}\partial_u X$, for a generic quantity $ X $.
These equations can be solved to give:
\begin{align}\label{eq:fastsolutions}
	\begin{split}
		L_S=&\dfrac{L_L^3}{(G_Nm)^2}\mathrm{Af}_{u_L}\Big(\int_0^{u_L}\!\!\!du\,\mathrm{Af}_{u}\Big(\dfrac{\partial\mathcal{R}}{\partial u}\Big)\Big) \; ,\\
		u_S=&-\mathrm{Af}_{u_L}\Big(\int_0^{u_L}\!\!\!du\,\Big[\dfrac{L_L^3}{(G_Nm)^2}\mathrm{Af}_{u}\Big(\dfrac{\partial\mathcal{R}}{\partial L}\Big)+\dfrac{3L_S}{L_L}\Big]\Big)\,.
	\end{split}
\end{align}
This is analogous to what obtained in Eq. \eqref{eq:psizero}, \eqref{eq:zetazero}, identifying $ h_1\mapsto -\partial \mathcal{R}/\partial L $ , $ g_1\mapsto\partial \mathcal{R} /\partial u$ and $\chi_0 H'/H\mapsto- 3L_S/L_L $.
A difference with respect to the toy model presented in Appendix \ref{app:toymodel} is the fact that now there are two small parameters, i.e. ratio of time-scales, which we have not factorized explicitly.
For convenience, in the following we will indicate $ \mathcal{F}=\mathrm{Af}_{u_L}(\mathcal{R}) $ and, using that partial derivatives commute with the average over $ u_L $, we will write $ \mathrm{Af}_{u_{L}}(\frac{\partial\mathcal{R}}{\partial X})=\frac{\partial \mathcal{F}}{\partial X} $.

Similar results will follow for the other contact elements and conjugate momenta.
At leading order, the equations of motion for the short-timescale variables, obtained through the near-identity transformation, will read
\begin{align} \label{eq:EOM_firstOrder}
	\begin{split}
		\partial_{u_L}{G}_S &= \dfrac{L_L^3}{(G_Nm)^2}\frac{\partial \mathcal{F}}{\partial \omega} \; , \quad \partial_{u_L}{\omega}_S = - \dfrac{L_L^3}{(G_Nm)^2}\frac{\partial \mathcal{F}}{\partial G}  \; , \\
		\partial_{u_L}{H}_S &= \dfrac{L_L^3}{(G_Nm)^2}\frac{\partial \mathcal{F}}{\partial \Omega}  \; , \quad \partial_{u_L}{\Omega}_S = - \dfrac{L_L^3}{(G_Nm)^2}\frac{\partial \mathcal{F}}{\partial H}  \; . 
	\end{split}
\end{align}
In order to convert derivatives with respect to canonical momenta in derivatives with respect to osculating contact elements, we use the formulas given in Appendix~\ref{app:LPE}.
As prescribed by the near-identity transformation approach, the quantities entering the right-hand side of Eqs.~\eqref{eq:EOM_firstOrder} are expressed only in terms of the long timescale variables $e_L$, $a_L$..., as derived in Eq. \eqref{eq:fasteom}. For simplicity, we will drop the $\hphantom{}_L$ subscript from now on, so that it will be understood that all osculating contact elements appearing in the perturbing function do not include quickly oscillating parts.
From the knowledge of these leading order oscillating parts of our variables, we can now compute the leading effect of back-reaction of the fast oscillations on the long time-scale dynamics, using the following procedure.
At the level of the Lagrangian, we can substitute the near-identity transformations and write the expression in~\eqref{eq:def_R} as
\begin{align}\label{eq:splitLag}
	\mathcal{L}=\mathcal{L}_L+\mathcal{L}_S \;,
\end{align}
the first term corresponding to the Lagrangian evaluated on the long time-scale variables only, while the second, $ \mathcal{L}_S $, corresponding to the remaining part, then to be averaged. For instance we have:
\begin{align}
	L\dot{u}=(L_L+L_S)(\dot{u}_L+\dot{u}_S)\;.
\end{align}
When we take the average of this quantity, the mixed terms $ L_L\dot{u}_S $ and $ L_S\dot{u}_L $, being exactly average free, will not contribute. Therefore, we only need to keep track of the two contributions  $ L_L\dot{u}_L $ in $ \mathcal{L}_L $ and  $ L_S\dot{u}_S $ in $ \mathcal{L}_S $. This means that, up to unimportant average-free terms, the long-timescale and short-timescale parts of the Lagrangian read
\begin{align}
    \mathcal{L}_L &= L_L\dot{u}_L+G_L\dot{\omega}_L+H_L\dot{\Omega}_L +  \dfrac{(G_Nm)^2}{2 L_L^2} +\langle\mathcal{R}\rangle \; , \\
	\mathcal{L}_S &= \dfrac{(G_Nm)^2}{2}\Big(\dfrac{1}{(L_L+L_S)^2}-\dfrac{1}{L_L^2}\Big)+L_S \dot u_S + G_S \dot \omega_S + H_S \dot \Omega_S+\sum_X \dfrac{\partial\mathcal{F}}{\partial X}X_S\;, \label{eq:LSwithXs}
\end{align}
where $ X $ represent contact elements and conjugate momenta, and we have used the splitting $\mathcal{R} = \mathcal{F} + \langle \mathcal{R} \rangle$ together with the fact that $\partial\langle \mathcal{R} \rangle / \partial X \; X_S$ is an average-free quantity. In this equation, it is understood as before that $\mathcal{R}$ and $\mathcal{F}$ are evaluated on the long-timescale variables only. We will approximate the first term as $ 3/2((G_Nm)^2/ L_L^4 )L_S^2\,$, since $ L_S $ is average-free and all the short timescale variables are suppressed either by $ \e^2 $ or by $ v^2 $.

Plugging back the equations of motion as well as the solutions obtained as in \eqref{eq:fastsolutions}, we find
\begin{align} \label{eq:LS}
	\mathcal{L}_S =& -\dfrac{L_L^3}{(G_N m)^2}\Bigg[\dfrac{\partial\mathcal{F}}{\partial u}\mathrm{Af}_{u_L}\Big(\int^{u_L}\!\!\!\!\!\! du\dfrac{\partial\mathcal{F}}{\partial L}\Big)+\dfrac{\partial\mathcal{F}}{\partial \omega}\mathrm{Af}_{u_L}\Big(\int^{u_L}_0 \!\!\!\!\!\! du\dfrac{\partial\mathcal{F}}{\partial G}\Big)+\dfrac{\partial\mathcal{F}}{\partial \Omega}\mathrm{Af}_{u_L}\Big(\int^{u_L}_0 \!\!\!\!\!\! du\dfrac{\partial\mathcal{F}}{\partial H}\Big)\Bigg]\nonumber\\
	& +\dfrac{3L_L^2}{(G_N m)^2}\Bigg[\dfrac{\mathcal{F}^2}{2}-\mathcal{F}^2-\dfrac{\partial\mathcal{F}}{\partial u}\mathrm{Af}_{u_L}\Big(\int^{u_L}\!\!\!\!\!\! du\mathcal{F}\Big)\Bigg]\;,
\end{align}
where in the second line, the first term comes from the expansion of $ 1/(L_L+L_S)^2 $, the second from substituting $ L_S\dot{u_S} $ and the third from substituting the solution found for $ u_S $ in Eq. \eqref{eq:fastsolutions}.
In order to simplify further this expression, it is useful to consider its average over $ u_L $. Considering the average allows us to perform integration by parts without having to deal with boundary terms, thanks to the fact that these are obtained subtracting the values of periodic functions at the endpoints $ u_L=0\,,\,2\pi $. Moreover, we can simplify Eq. \eqref{eq:LS} using that $ \langle \mathrm{Af}(M)\mathrm{Af}(N)\rangle=\langle\mathrm{Af}(M)N\rangle $ for any arbitrary functions $M$, $N$. Thus we find: 
\begin{align} \label{eq:LSaveraged}
	\langle\mathcal{L}_S \rangle=& \dfrac{1}{2\pi}\int_{0}^{2\pi}\!\!\!\!\!\! du_L\Bigg[\dfrac{L_L^3}{(G_N m)^2}\Bigg(\dfrac{\partial\mathcal{F}}{\partial L}\mathcal{F}+\dfrac{\partial\mathcal{F}}{\partial  G}\int^{u_L}_0 \!\!\!\!\!\! du\dfrac{\partial\mathcal{F}}{\partial \omega}+\dfrac{\partial\mathcal{F}}{\partial H}\int^{u_L}_0 \!\!\!\!\!\! du\dfrac{\partial\mathcal{F}}{\partial \Omega}\Bigg) +\dfrac{3L_L^2}{2(G_N m)^2}\mathcal{F}^2\Bigg]\;.
\end{align}


In the next Subsection, we will compute the quadrupolar post-Newtonian cross-terms given by this procedure. 
Using the same procedure, we can also obtain the so-called "quadrupole-squared" terms studied in~\cite{Will:2020tri}, which are a purely Newtonian contribution of post-adiabatic corrections. As a proof-of-concept, we use our procedure to compute such terms in Appendix~\ref{app:quad_squared} and show that they give back the exact same equations as the ones displayed in~\cite{Will:2020tri}.

\subsection{1PN quadrupolar cross-terms} \label{sec:1PNquadCrossTerms}

We now need the expression of the perturbing function $\mathcal{R}$, which contains both post-Newtonian and quadrupolar terms. They are obtained by expanding the Lagrangian~\eqref{eq:L1PN_unexpanded} in the center-of-mass frame. In doing so, one obtains a quadrupolar part $\mathcal{R}_\mathrm{quad}$, of order $\varepsilon^2$ within our power-counting rules, and a post-Newtonian part $\mathcal{R}_\mathrm{1PN}$. The latter is itself composed of two terms $\mathcal{R}_\mathrm{1PN} = \mathcal{R}_{v^2}+\mathcal{R}_{v^2 \varepsilon^{1/2}}$ scaling differently: the first one is of order $v^2$ and corresponds to the usual EIH Lagrangian in the center-of-mass frame, and the second one is linear in $V_\mathrm{CM}$ and thus of order $v^2 \varepsilon^{1/2}$. Higher-order terms in the $\varepsilon$ expansion can be safely neglected for the precision we aim to.

Now, a helpful simplification comes directly from using the relative coordinates of the inner binary center of mass as explained in Section~\ref{sec:CM}. Indeed, once we express the Lagrangian in terms of these coordinates, the contributions of order $v^2 \varepsilon^{1/2}$ precisely cancel each other, so that they do not contribute to cross-terms. This cancellation is made possible by the fact that the change of reference frame mixes different orders of the expansion. Had we defined the contact elements through the three-body center-of-mass relative coordinates in Eq.~\eqref{eq:x1x2_CM}, this term would have been non-zero and would have lead to cumbersome formulas which would have prevented us from performing the matching of the inner binary to a point-particle as we did in Section~\ref{sec:matching}.
We now give the expression of the different perturbing functions in terms of $r$ and $v$ as
\begin{align}\label{eq:R_1PN_quad}
	\begin{split}
		\mathcal{R}_{v^2} &= \frac{1}{8} v^4 (1-3 \nu) + \frac{G_N m}{2r} \left( (3+\nu)v^2 + \nu (\mathbf{v} \cdot \mathbf{n})^2 - \frac{G_N m}{r} \right) \; , \\
		\mathcal{R}_\mathrm{quad} &= -\frac{1}{2} r^i r^j \partial_i \partial_j \tilde \phi \; ,
	\end{split}
\end{align}
where we recall that $\tilde \phi$ is an arbitrary external field, $X_A = m_A/m$ and $\nu = X_1 X_2$.
Substitution of near identity transformation in these expressions will produce 1PN quadrupolar terms, i.e of order $v^2 \varepsilon^2$.



Using the expression of the osculating elements, we find
\begin{align} 
	\mathcal R_{v^2}  &= \frac{G_N^2 m^2}{2a^2 (1-e \cos \eta)^2} \bigg[ \frac{1-3 \nu}{4} (1+e \cos \eta)^2 + (3 + \nu)(1+e \cos \eta) + \nu e^2 \frac{\sin^2 \eta}{1 - e\cos \eta} -1  \bigg]\label{eq:R1PN_Rquad_1} \; , \\
	\mathcal{R}_\mathrm{quad} &= - \frac{a^2}{2} \big( (\cos \eta -e) \; \alpha^i + \sqrt{1-e^2} \sin \eta \; \beta^i \big) \big( (\cos \eta -e) \; \alpha^j + \sqrt{1-e^2} \sin \eta \; \beta^j \big) \partial_i \partial_j \tilde \phi \;.\label{eq:R1PN_Rquad_2}
\end{align}

We can now compute the cross-terms following Eq.~\eqref{eq:LSaveraged}.
The Lagrangian displayed in Eq.~\eqref{eq:LSaveraged} contains cross-terms of order $v^2 \varepsilon^2$. It also contains 2PN ($v^4$) and quadrupole-squared ($\varepsilon^4$) terms, which we will ignore since we limit ourselves to 1PN quadrupolar order. 
Cross-terms of order $v^2 \varepsilon^2$ will not receive contributions from the term containing both $ H $ and $ \Omega $ derivatives, since $ \mathcal{R}_{v^2} $ does not depend on neither of these two variables. 
Computations are performed using the \textit{Mathematica} software, giving the full expression of cross-terms in terms of osculating elements:

\begin{mybox}
	\begin{align}\label{eq:L1_crossTerms}
		\begin{split}
			\langle\mathcal{L}_S\rangle &= \frac{G_N m a}{8 (1-e^2 +\sqrt{1-e^2})} \bigg\lbrace [2(9+7 \sqrt{1-e^2}) + e^2(51+11 \sqrt{1-e^2}) + 16 e^4] \alpha^i \alpha^j \\
			&+ [2(7+9 \sqrt{1-e^2}) + e^2(29+17 \sqrt{1-e^2}) -8 e^4] \beta^i \beta^j  \bigg\rbrace \partial_i \partial_j \tilde \phi \;.\\
		\end{split}
	\end{align}
\end{mybox}
This term will be part of the effective Lagrangian for the long time-scale dynamics:
\begin{align}\label{eq:Leff}
	\mathcal{L}_{eff}=\langle\mathcal{L}_L\rangle+\langle\mathcal{L}_S\rangle=L_L\dot{u}_L+G_L\dot{\omega}_L+H_L\dot{\Omega}_L+\langle\mathcal{R}\rangle+\langle\mathcal{L}_S\rangle \;.
\end{align}
This effective Lagrangian contains the term $ \mathcal{L}_{\mathrm{quad},12}^{\leq v^2\varepsilon^{5/2}} $ computed in Section \ref{sec:average}, but it also includes the terms at lower order in the multipole expansion.

Note that other cross-terms, besides those contained in $ \langle\mathcal{L}_S\rangle $ and those coming from terms of order quadrupole-1PN in $ \mathcal{R} $, might come in principle if the change of variables from the mean anomaly to the eccentric anomaly, discussed in Appendix \ref{app:toymodel}, was to differ with respect to the change of variables derived from the Kepler equation \eqref{eq:kepler}. As discussed, we change variable from the long time-scale part of the mean anomaly, to the long time-scale part of the eccentric anomaly, defined so as to be related to the former by the Kepler equation. This removes any potential cross term contribution due to the change of variables in the integration.

Finally, note also that to this order cross terms in the effective Lagrangian \eqref{eq:Leff} will not get contributions due to subleading oscillating parts of the near-identity transformations (e.g. short time-scale functions with amplitude of order quadrupole-1PN, similar to $x_S^{(1)}$ in the notation of Appendix \ref{app:toymodel}), since these will be average free and will have to multiply terms of $ \mathcal{L}_L $, leading to average-free contributions, if any.

\section{Lagrange Planetary Equations} \label{app:LPE}

The Lagrange Planetary Equations (LPE) express the time-derivatives of all osculating (contact) elements in terms of derivatives of the Hamiltonian~\cite{relativistic_celestial_mechanics}. They can be obtained as the equations of motion stemming from the Lagrangian~\eqref{eq:def_R}. The derivatives of $\mathcal{R}$ with respect to canonical momenta can be transformed in derivatives with respect to planetary elements by inverting the relations~\eqref{eq:def_L_G_H}:
\begin{equation}
a = \frac{L^2}{Gm} \; , \quad e = \sqrt{1 - \frac{G^2}{L^2}} \; , \quad \cos \iota = \frac{H}{G} \; ,
\end{equation}
Thus, one has
\begin{align}\label{eq:derivatives_momentum}
\frac{\partial \mathcal{R}}{\partial L} &= 2 \sqrt{\frac{a}{G_N m}} \frac{\partial \mathcal{R}}{\partial a} + \frac{1-e^2}{e \sqrt{G_N ma}} \frac{\partial \mathcal{R}}{\partial e} \; , \\
 \frac{\partial \mathcal{R}}{\partial G} &= - \frac{\sqrt{1-e^2}}{e \sqrt{G_N ma}} \frac{\partial \mathcal{R}}{\partial e} + \frac{\cos \iota}{\sqrt{G_Nma(1-e^2)} \sin \iota} \frac{\partial \mathcal{R}}{\partial \iota} \; , \\
 \frac{\partial \mathcal{R}}{\partial H} &= -\frac{1}{\sqrt{G_Nma(1-e^2)} \sin \iota} \frac{\partial \mathcal{R}}{\partial \iota} \; .
\end{align}
In these equations, the derivatives with respect to the osculating elements are taken holding all other elements fixed. However, we should be careful because the eccentric anomaly $\eta$ depends on both the eccentricity and the semimajor axis through the equation $\eta - e \sin \eta = \sqrt{G_N m/a^3} t + \sigma$. Consequently,
\begin{equation}
\left. \frac{\partial \eta}{\partial e} \right\vert_{\sigma, a \; \mathrm{ fixed}} = \frac{\sin \eta}{1-e\cos \eta} \; , \quad \left. \frac{\partial \eta}{\partial a} \right\vert_{\sigma, e \; \mathrm{ fixed}} = - \frac{3 n t}{2a(1-e\cos \eta)} \; .
\end{equation}
Thus,
\begin{equation}
\frac{\partial \mathcal{R}}{\partial e} = \left. \frac{\partial \mathcal{R}}{\partial e} \right\vert_\mathrm{\eta \; fixed} +  \frac{\sin \eta}{1-e\cos \eta} \frac{\partial \mathcal{R}}{\partial \eta} \; , \quad \frac{\partial \mathcal{R}}{\partial a} = \left. \frac{\partial \mathcal{R}}{\partial a} \right\vert_\mathrm{\eta \; fixed} - \frac{3 n t}{2a(1-e\cos \eta)} \frac{\partial \mathcal{R}}{\partial \eta} \; .
\end{equation}

Apart from this subtelty, one can now easily derive the LPE from the Lagrangian~\eqref{eq:def_R}, and they read
\begin{align}
\dot a &= \sqrt{\frac{4a}{G_N m}} \frac{\partial  \mathcal{R}}{\partial u} \; , \label{eq:dot_a} \\
\dot e &= - \sqrt{\frac{1-e^2}{G_N m a e^2}} \frac{\partial  \mathcal{R}}{\partial \omega} + \frac{1-e^2}{\sqrt{G_N m a}e} \frac{\partial  \mathcal{R}}{\partial u} \; , \label{eq:dot_e} \\
\dot \iota& = - \frac{1}{\sqrt{G_N m a(1-e^2)} \sin \iota} \frac{\partial  \mathcal{R}}{\partial \Omega}
+ \frac{\cos \iota}{\sqrt{G_N m a(1-e^2)} \sin \iota} \frac{\partial  \mathcal{R}}{\partial \omega} \; , \\
\dot u &= \sqrt{\frac{G_N m}{a^3}} - \sqrt{\frac{4a}{G_N m}} \frac{\partial \mathcal{R}}{\partial a} - \frac{1-e^2}{\sqrt{G_N m a}e} \frac{\partial  \mathcal{R}}{\partial e} \; , \label{eq:dot_u} \\
\dot \omega &= \sqrt{\frac{1-e^2}{G_N m a e^2}} \frac{\partial  \mathcal{R}}{\partial e} - \frac{\cos \iota}{ \sqrt{G_N m a(1-e^2)} \sin \iota} \frac{\partial \mathcal{R}}{\partial \iota} \; , \label{eq:dot_omega} \\
\dot \Omega &=  \frac{1}{\sqrt{G_N m a(1-e^2)} \sin \iota} \frac{\partial \mathcal{R}}{\partial \iota} \; . \label{eq:dot_Omega}
\end{align}
The averaged equations can be obtained as explained in Appendix \ref{app:toymodel}, through a near-identity transformation. The LPE for the outer orbit are obtained in the very same way, replacing all inner quantities with outer ones. However, one has to be careful to replace the mass $m$ with the sum $m_3 + \mathcal{E}$, with $\mathcal{E} = m - G_N m/2a$, as was emphasized below Eq.~\eqref{eq:totLagrangian}.

Finally, let us make two technical remarks. First, in order to express derivatives of the quadrupolar Lagrangian in Eq.~\eqref{eq:R1PN_Rquad_2} with respect to angles, we use the following derivatives of the basis vectors:
\begin{align}
\begin{split} \label{eq:derivatives_basis}
\frac{\partial \alpha^i}{\partial \omega} &= \beta^i \; , \quad \frac{\partial \beta^i}{\partial \omega} = - \alpha^i \; , \\
\frac{\partial \alpha^i}{\partial \Omega} &= - \cos \omega \sin \iota \; \gamma^i + \cos \iota \; \beta^i \; , \quad \frac{\partial \beta^i}{\partial \Omega} = - \cos \iota \; \alpha^i + \sin \iota \sin \omega \; \gamma^i \; , \\
\frac{\partial \alpha^i}{\partial \iota} &= \sin \omega \; \gamma^i \; , \quad \frac{\partial \beta^i}{\partial \iota} =  \cos \omega \; \gamma^i  \; .
\end{split}
\end{align}

Second, we will use (spatial) gauge-invariance in order to simplify the computations as much as possible. Indeed, we know that the Lagrangian is a rotation-invariant quantity. Thus, when carrying out the average of the first-order Lagrangian obtained by integrating out the high-energy modes, we use a particular gauge choice for the angles: $\iota = \pi/2$, $\omega = 0$. The resulting averaged Lagrangian will be gauge-invariant provided we express it only in terms of the basis vectors $\bm \alpha$, $\bm \beta$, $\bm \gamma$. In this gauge, the derivatives written above simplify greatly:
\begin{align}
\frac{\partial \alpha^i}{\partial \omega} &= \beta^i \; , \quad \frac{\partial \beta^i}{\partial \omega} = - \alpha^i \; , \quad \frac{\partial \alpha^i}{\partial \Omega} = -\gamma^i \; , \quad \frac{\partial \beta^i}{\partial \Omega} = 0 \; , \quad \frac{\partial \alpha^i}{\partial \iota} = 0 \; , \quad \frac{\partial \beta^i}{\partial \iota} = \gamma^i \; , \\
\frac{\partial \mathcal{R}}{\partial L} &= 2 \sqrt{\frac{a}{Gm}} \frac{\partial \mathcal{R}}{\partial a} + \frac{1-e^2}{e \sqrt{Gma}} \frac{\partial \mathcal{R}}{\partial e} \; , \quad \frac{\partial \mathcal{R}}{\partial G} = - \frac{\sqrt{1-e^2}}{e \sqrt{Gma}} \frac{\partial \mathcal{R}}{\partial e}  \; , \quad \frac{\partial \mathcal{R}}{\partial H} = - \frac{1}{\sqrt{Gma(1-e^2)}} \frac{\partial \mathcal{R}}{\partial \iota} \; .
\end{align}

To give further support to the validity of this simplification we refer the reader to Appendix~\ref{app:quad_squared}, where we explicitly derive that the spurious dependence on angles contained in the derivatives~\eqref{eq:derivatives_basis} exactly cancel when averaging the high-energy modes in order to obtain the quadrupole-squared terms.

\section{From the three-body center-of-mass frame to the inner binary rest frame} \label{app:CM}

In this Appendix we derive the explicit relation between the absolute coordinates $(\bm y_1, \bm y_2)$ of the inner binary and the relative distance $\bm r'$ defined in the inner binary rest frame, which is the natural quantity in terms of which one can express osculating elements, as discussed in Section~\ref{sec:CM}.
Let us begin by recalling the post-Newtonian definition of the center-of-mass of the relativistic two-body system composed by the inner binary, given to 1PN order by:
\begin{align} \label{eq:CM_1PN}
\begin{split}
E \mathbf{Y}_\mathrm{CM} &= E_1 \mathbf{y}_1 + E_2 \mathbf{y}_2 \; , \\  E = E_1 + E_2 \quad ,\quad  E_A &= m_A + \frac{1}{2} m_A v_A^2 - \frac{G_N m_1 m_2}{2 r} \;,
\end{split}
\end{align}
for $ A=1,2 $.
In the three-body rest frame $ \mathsf{R} $, the relation $\bm r = \bm y_1 - \bm y_2$ leads to
\begin{equation}\label{eq:x1x2_CM}
\mathbf{y}_1 =  \mathbf{Y}_\mathrm{CM} + (X_2 + \delta) \mathbf{r} \; , \quad \mathbf{y}_2 =  \mathbf{Y}_\mathrm{CM} + (-X_1 + \delta) \mathbf{r} \; ,
\end{equation}
where we have defined
\begin{align} \label{eq:def_CM_PN}
\begin{split}
X_A &= \frac{m_A}{m} \; , \quad m = m_1 + m_2 \; , \quad \mu =  \frac{m_1 m_2}{m} \; , \quad \nu = \frac{\mu}{m} \; , \\
\delta &= - \nu \mathbf{V}_\mathrm{CM} \cdot \mathbf{v} + \nu (X_1 - X_2) \left( \frac{v^2}{2} - \frac{G_N m}{2r} \right).
\end{split}
\end{align}
However, the relative distance $ \boldsymbol{r} $ in the three-body rest frame $ \mathsf{R} $ cannot be expressed in terms of the \textit{intrinsic} contact elements using Eq. \eqref{eq:ParamContactElements}. Rather, it will be the relative distance in the inner binary rest frame $ \mathsf{R}' $, $ \boldsymbol{r}'=\boldsymbol{y}_1'-\boldsymbol{y}_2' $, and the respective momentum to be related to these convenient variables by Eq. \eqref{eq:ParamContactElements}.
\begin{figure}
\centering
\includegraphics[width=0.3\columnwidth]{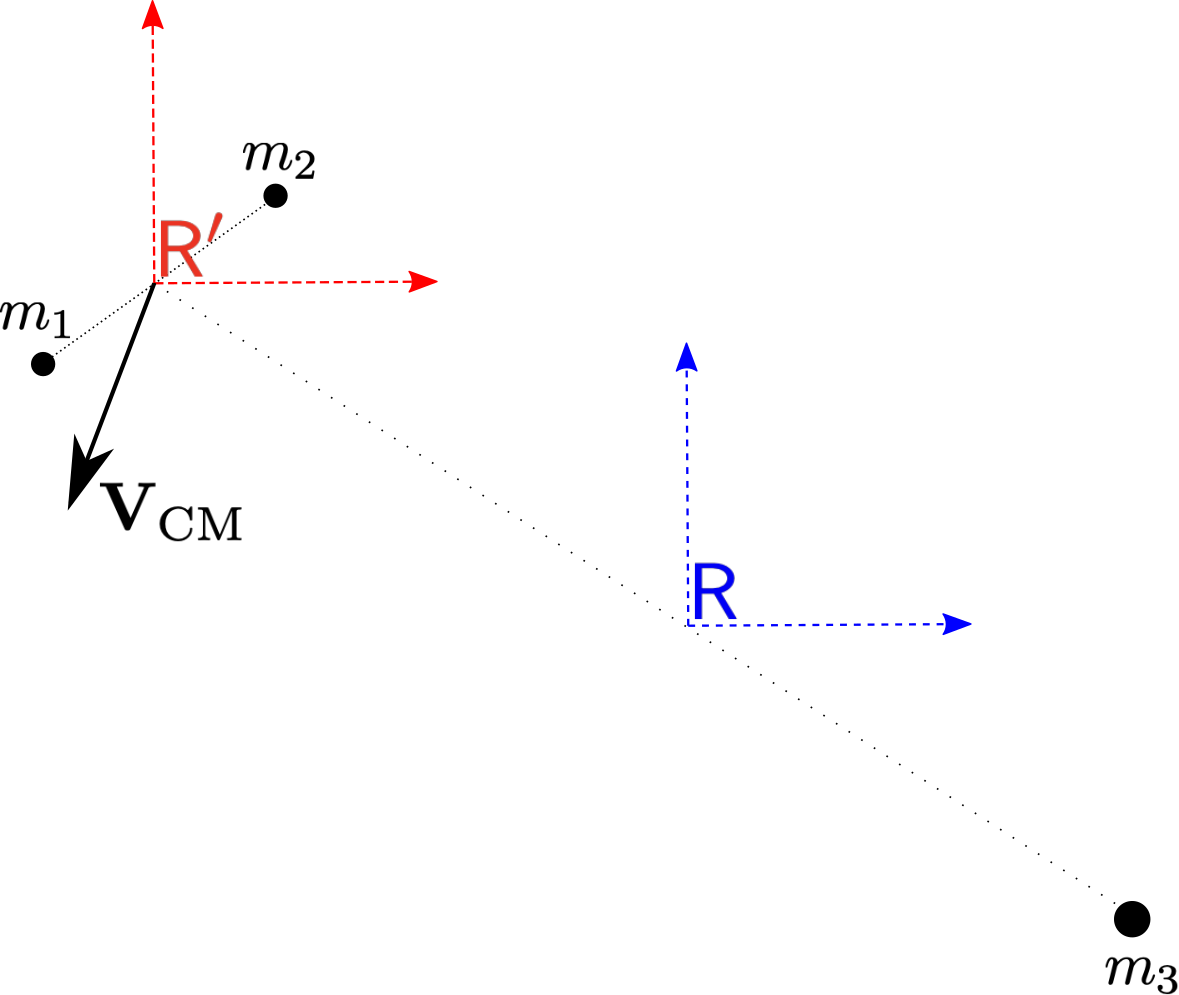}
\caption{Illustration of the change of referential~\eqref{eq:lorentz_transfo}: the rest frame of the inner binary $\mathsf{R}'$ is obtained from the total center-of-mass frame $\mathsf{R}$ by translating it by $\bm X_\mathrm{CM}$ and boosting it by $\bm V_\mathrm{CM}$. }
\label{fig:change_ref}
\end{figure}
For this reason, we want to express the relative coordinates in the rest frame $\mathsf{R}$ in terms of those defined in $\mathsf{R}'$. Concretely, around a time $ t_0 $, the two coordinate systems are related by the following transformation:
\begin{equation} \label{eq:lorentz_transfo}
\begin{pmatrix}
t' - t_0 \\
\bm y'
\end{pmatrix}
 = \mathcal{B} \begin{pmatrix}
t - t_0 \\
\bm y - \bm Y_\mathrm{CM}(t_0)
\end{pmatrix} \; ,
\end{equation}
where $\mathcal{B}$ represents a Lorentz boost of velocity $V_\mathrm{CM}(t_0)$. To 1PN order, the coordinates of the two inner bodies $\bm y_A'$ (with $A=1,2$) in the frame $ \mathsf{R}' $ are related to those of the three-body rest frame $ \mathsf{R} $ by:
\begin{equation}
\bm y_A'(t') = \bm y_A(t') - \bm Y_\mathrm{CM} - \bm V_\mathrm{CM}(t'-t_0) + \bm V_\mathrm{CM} \cdot \big(\bm y_A(t') - \bm Y_\mathrm{CM} \big) \bigg( \bm v_A(t') - \frac{\bm V_\mathrm{CM}}{2} \bigg) \; ,
\end{equation}
where in the last equation it is understood that $\bm Y_\mathrm{CM}$ and $\bm V_\mathrm{CM}$ are evaluated at $t_0$. 
Inserting this relation in the center-of-mass definition~\eqref{eq:CM_1PN} to 1PN order, we obtain
\begin{equation}
E_1 \bm y_1' + E_2 \bm y_2' =  \big( \bm V_\mathrm{CM} \cdot \bm r' \big) \bm p' \; ,
\end{equation}
where we have substituted $ \mu\boldsymbol{v}'\simeq \boldsymbol{p}' $. From this, we find that the relation between absolute coordinates $ \boldsymbol{y}_A' $ and relative coordinates $ \boldsymbol{r}'\,,\,\boldsymbol{v}' $ in rest frame of the inner binary is:
\begin{align}\label{eq:abstorel}
\begin{split}
\bm y_1' &= \big( X_2+\delta\big) \bm r' + \big( \bm V_\mathrm{CM} \cdot \bm r' \big) \bm p'/m \; , \\
\bm y_2' &= \big( -X_1+\delta\big) \bm r' + \big( \bm V_\mathrm{CM} \cdot \bm r' \big) \bm p'/m\;,
 \end{split}
\end{align}
where $\delta$ is the 1PN quantity already defined in Eq.~\eqref{eq:def_CM_PN} (the use of primed or unprimed quantities in $\delta$ does not matter since the difference would be of 2PN order), and it is understood that $\bm y_1'$, $\bm y_2'$, $\bm r'$ and $\bm v'$ are evaluated at the same time $t'$.

This result allows to express the three-body frame coordinates $ \boldsymbol{y}_A $ in terms of the relative coordinates in the rest frame of the inner binary, $ \boldsymbol{r}'\,,\,\boldsymbol{p}' $ \footnote{Note that $ \boldsymbol{p}' $ is defined through the relative velocity $ \boldsymbol{v}' $, the $ t'$ derivative of $ \boldsymbol{r}'
$.}, which can be in turn expressed in terms of \textit{intrinsic} contact elements using Eq. \eqref{eq:ParamContactElements}\footnote{Although Eq. \eqref{eq:ParamContactElements} could be used to define osculating elements whatever the frame, now we will apply it to the relative distance and momentum in the binary rest frame, which we are calling $ \boldsymbol{r} '$ and $ \boldsymbol{p} '$.}. We obtain:
\begin{align}\label{eq:x1x2_CM'_app}
\begin{split}
\bm{y}_1 &=  \mathbf{Y}_\mathrm{CM} + \mathbf{V}_\mathrm{CM} (t-t_0) + (X_2 + \delta) \mathbf{r}' + X_2 \big( \bm V_\mathrm{CM} \cdot \bm r' \big) \bigg[ \big(X_1-X_2\big) \frac{\bm p'}{\mu} - \frac{\bm V_\mathrm{CM}}{2} \bigg] \; , \\
 \quad \bm{y}_2 &=  \mathbf{Y}_\mathrm{CM} + \mathbf{V}_\mathrm{CM} (t-t_0) + (-X_1 + \delta) \mathbf{r}' - X_1 \big( \bm V_\mathrm{CM} \cdot \bm r' \big) \bigg[ \big(X_1-X_2\big) \frac{\bm p'}{\mu} - \frac{\bm V_\mathrm{CM}}{2} \bigg] \; ,
 \end{split}
\end{align}
where $\bm r'$ and $\bm p'$ are evaluated at the time $t$, while $\mathbf{Y}_\mathrm{CM}$ and $\mathbf{V}_\mathrm{CM}$ are evaluated at $t_0$. 

In particular, this result implies the following relation between $\bm r = \bm y_1 - \bm y_2$ and $\bm r'$:
\begin{equation}
	\bm r = \bm r' - \big( \bm V_\mathrm{CM} \cdot \bm r' \big) \bigg[ \frac{\bm V_\mathrm{CM}}{2} + \big( X_2-X_1 \big) \frac{\bm p'}{\mu} \bigg] \; .
\end{equation}
The final step is just to evaluate the above expressions \eqref{eq:x1x2_CM'_app} and their time-derivatives at $t=t_0$ and to express $ \bm r' \,,\,\bm p' $ in terms of the contact elements as in Eq. \eqref{eq:ParamContactElements}.
To avoid clutter, in the main text and the rest of the Appendices we will suppress the primed label on $\bm r'$, $\bm p'$.

\section{Quadrupole-squared terms} \label{app:quad_squared}

In this Appendix we use the methodology developed in Appendix~\ref{app:beyond_adiab} to compute the so-called quadrupole-squared terms presented in~\cite{Will:2020tri}. These purely Newtonian contributions to the evolution of a \hie system come from deviations to the adiabatic approximation as well as backreaction of quickly oscillating modes when averaging quadrupolar terms. The magnitude of these contributions can be greater than octupole order terms so that they could induce interesting deviations to the Kozai-Lidov mechanism.

As stated in Appendix~\ref{app:beyond_adiab}, corrections to the leading order averaging could come at the level of either the inner binary or the outer binary orbital motion, because of short-timescale fluctuations of the form $X=X_L+X_S$ where $X$ is any osculating element or canonical momentum of the inner or outer binary. More precisely, there are four kind of short-timescale fluctuations of osculating elements to consider: (\textit{i}) fluctuations of the inner planetary elements on an inner binary timescale; (\textit{ii}) fluctuations of the outer planetary elements on an inner binary timescale; (\textit{iii}) fluctuations of the inner planetary elements on an outer binary timescale; (\textit{iv}) fluctuations of the outer planetary elements on an outer binary timescale. In all these cases, a generalization of the methodology developed in Appendix~\ref{sec:LS} shows that the cross-terms in $L_S$~\eqref{eq:LSaveraged} scale as (see in particular Eq.~\eqref{eq:LSwithXs} and Eq.~\eqref{eq:EOM_firstOrder}):
\begin{align}
(i) \; & \frac{\mathcal{F}^2}{L} \frac{L^3}{G_N^2 m^2} \sim \mathcal{F}^2 \frac{a}{G_N m} \; , \\
(ii) \; & \frac{\mathcal{F}^2}{L_3} \frac{L^3}{G_N^2 m^2} \sim \mathcal{F}^2 \frac{a}{G_N m} \varepsilon^{1/2} \; , \\
(iii) \; & \frac{\mathcal{F}^2}{L} \frac{L_3^3}{G_N^2 m^2} \sim \mathcal{F}^2 \frac{a}{G_N m} \varepsilon^{-3/2} \; , \\
(iv) \; & \frac{\mathcal{F}^2}{L_3} \frac{L_3^3}{G_N^2 m^2} \sim \mathcal{F}^2 \frac{a}{G_N m} \varepsilon^{-1} \; .
\end{align}

This makes it clear that, in general, case \textit{(iii)} gives the largest contribution of the cross-terms (see also the interesting discussion in Appendix B of Ref.~\cite{Will:2020tri}). This scaling is perfectly valid for the quadrupole-squared terms which we want to compute in this Appendix; instead, in the previous case of quadrupole-1PN cross-terms, there is an additional suppression on the outer binary timescale which explains that the dominant contribution to cross-terms comes from case \textit{(i)} as computed in Appendix~\ref{app:beyond_adiab}. This additional suppression comes from the fact that the PN terms in the function $\mathcal{F}$ come with an additional $\varepsilon^2$ multiplicative factor for cases (\textit{iii}) and (\textit{iv}) --- i.e. when averaging over the outer binary timescale. To see this, remark that the 1PN perturbing function can be schematically written as: $\mathcal{R}_\mathrm{1PN} = \mathcal{R}_{v^2} + \mathcal{R}_{v^2 \varepsilon^{1/2}} + \mathcal{R}_{v^2 \varepsilon} + \mathcal{R}_{v^2 \varepsilon^{3/2}} + \mathcal{R}_{v^2 \varepsilon^{2}} $. This splitting, as already emphasized in Appendix~\ref{sec:1PNquadCrossTerms}, comes from introducing the center-of-mass coordinates~\eqref{eq:x1x2_CM'} in the EIH Lagrangian~\eqref{eq:LEIH}. With our center-of-mass choice, both $\mathcal{R}_{v^2 \varepsilon^{1/2}}$, $\mathcal{R}_{v^2 \varepsilon}$ and $\mathcal{R}_{v^2 \varepsilon^{3/2}}$ vanish after averaging on the inner binary timescale. Thus, the only term featuring a non-trivial dependence on the outer binary timescale is $\mathcal{R}_{v^2 \varepsilon^2}$, since $\mathcal{R}_{v^2}$ is just the standard EIH Lagrangian for the inner binary and does not depend on the outer binary period. This proves the additional $\varepsilon^2$ suppression of outer binary averages, justifying the use of case \textit{(i)} in Appendix~\ref{app:beyond_adiab}.



Let us now derive the quadrupole-squared terms using the scaling \textit{(iii)}. Since we are only interested in deviations from adiabaticity in the \textit{outer} average, we only perform leading order averaging for the inner binary timescale.
Having removed the dependence of the Lagrangian on the mean anomaly of the inner orbit, the semimajor axis of the inner orbit will remain constant. This means that this variable will not have short oscillations on time-scales of the order of the period of the outer binary. The other variables instead will have modes that evolve during the period of the outer orbit and modes that evolve on much longer time-scales. Therefore we will generally write $ X=X_{\tilde{L}}+X_{\tilde{S}} $, indicating a near identity transformation having the period of the outer binary as reference time-scale. As indicated by the scaling \textit{(iii)}, we only consider such a decomposition for $X$ being an inner osculating element, because fluctuations of outer osculating elements are suppressed by an additional $\varepsilon^{1/2}$ factor.


At this point we can already check that the quadrupole-square cross terms Lagrangian $\mathcal{L}_{\tilde{S}}$ scales as
\begin{equation}
\frac{\mathcal{L}_{\tilde{S}}}{\mathcal{L}_{\tilde{L}}} \sim \frac{n_3}{n} \sim \sqrt{\frac{M a^3}{m a_3^3}}\;.
\end{equation}
Thus, despite the quadrupole-squared terms are formally a $\varepsilon^{3/2}$ perturbation to the quadrupole (i.e between octupole and hexadecapole), they could be enhanced to a greater magnitude than the octupole if the ratio $M/m$ is large.

To proceed with the computation, we will indicate the quadrupolar part of the perturbing function in Eq.~\eqref{eq:Lquad_final_before_avg} to Newtonian order as:
\begin{align}
	\mathcal{R}_{Q3}= \frac{3 G_N m_3}{2 \mu} Q^{ij} \frac{N_i N_j}{R^3} \; , \quad Q^{ij} = \frac{\mu a^2}{2} \bigg( (1+4e^2) \alpha^i \alpha^j + (1-e^2) \beta^i \beta^j - \frac{2+3e^2}{3} \delta^{ij} \bigg) \; ,
\end{align}
and its average free part, taken with respect to the long time-scale part of the outer mean anomaly $ u_{3\,\tilde{L}} $,  as:
\begin{align}
	\mathcal{F}_{Q3}=\mathcal{R}_{Q3}-\frac{1}{2 \pi}\int^{2\pi}_0 du_{3\,\tilde{L}}\mathcal{R}_{Q3}\;,
\end{align}
where we recall that $\mathbf{R}$ is the distance vector of the outer orbit, $R = \vert \bm R \vert$ and $\bm N = \bm R / R$.
With this notation, using the equations of motion and expanding the perturbing function, we obtain:
\begin{align}
	\mathcal{L}_{\tilde{S}}&=L_{\tilde{S}}\dot{u}_{\tilde{S}}+G_{\tilde{S}}\dot{\omega}_{\tilde{S}}+H_{\tilde{S}}\dot{\Omega}_{\tilde{S}}+\sum_{X}\dfrac{\partial \mathcal{F}_{Q3}}{\partial X}X_{\tilde{S}}
	\\\nonumber
	&=\mathrm{Af}\Big(\int^{u_{3\,\tilde{L}}}ds\dfrac{\partial\mathcal{F}_{Q3}}{\partial G}\Big)\dfrac{\partial\mathcal{F}_{Q3}}{\partial \omega}+\mathrm{Af}\Big(\int^{u_{3\,\tilde{L}}}ds\dfrac{\partial\mathcal{F}_{Q3}}{\partial H}\Big)\dfrac{\partial\mathcal{F}_{Q3}}{\partial \Omega}\;.
\end{align}
Then, performing the averages with respect to the long time-scale part of the outer mean anomaly, we find:
\begin{equation} \label{eq:L1Mean}
\left\langle \mathcal{L}_S \right\rangle = - \frac{9 G_N^2 m_3^2}{4 \mu^2} \left( \frac{\partial Q^{ij}}{\partial \omega} \frac{\partial Q^{kl}}{\partial G} + \frac{\partial Q^{ij}}{\partial \Omega} \frac{\partial Q^{kl}}{\partial H} \right) \mathcal{A}_{ij;kl} \equiv  \mathcal{B}^{ij;kl} \mathcal{A}_{ij;kl} \; ,
\end{equation}
where the tensor $\mathcal{A}_{ij;kl}$ is a sort of variance given by
\begin{equation}
\mathcal{A}_{ij;kl} = \left\langle \left[ \frac{N_i N_j}{R^3} - \left\langle \frac{N_i N_j}{R^3} \right\rangle \right] \times  \int \mathrm{d} t\left[ \; \frac{N_k N_l}{R^3} - \left\langle \frac{N_k N_l}{R^3} \right\rangle \right] \right\rangle \; .
\end{equation}
Note that the choice of a constant in the integration does not matter since it multiplies a zero-mean term. $\mathcal{A}_{ij;kl}$ is symmetric in the $(i,j)$ as well as in the $(k,l)$ indices, and it is antisymmetric by exchange of the pairs $(i,j)$ and $(k,l)$ (by an integration by parts). There now just remains to compute the derivatives in Eq.~\eqref{eq:L1Mean}. We will use the derivatives given in Appendix~\ref{app:LPE}, Eqs.~\eqref{eq:derivatives_momentum} and~\eqref{eq:derivatives_basis}. These derivatives generically depend on the angles $\Omega$, $\omega$ and $\iota$. However, such angles cannot remain in the final result for $\left\langle \mathcal{L}_S \right\rangle$, since the contraction in the spatial indices needs to transform correctly under a rotation. This is analogous to the 'background field method' in EFTs: if we integrate out some fluctuating field in some given gauge, then the computational steps can be gauge-dependent but the final result should be gauge-invariant since it is expressed as a (gauge-invariant) long-wavelength Lagrangian. The cancellation of the angular dependence will be a very non-trivial check of our computation. In the more complicated computation of the PN quadrupolar cross-terms, instead, we have chosen to reverse the argument and choose a particular gauge (i.e, a particular value for $\Omega$, $\omega$ and $\iota$) to simplify the calculations, as explained at the end of Appendix~\ref{app:LPE}.


We now go on for the final computation. In the $\mathcal{B}^{ij;kl}$ tensor, the dependence on the angle $\iota$ nicely factors out:
\begin{align}
\begin{split}
\frac{1}{ \sin \iota} \frac{\partial Q^{kl}}{\partial \iota} \left(  \cos \iota \frac{\partial Q^{ij}}{\partial \omega} - \frac{\partial Q^{ij}}{\partial \Omega} \right) &= \frac{\mu^2 a^4}{4} \bigg[ \big(1+4e^2\big) \cos \omega \big( \alpha^i \gamma^j + \mathrm{sym} \big) - \big(1-e^2\big) \sin \omega \big( \beta^i \gamma^j + \mathrm{sym} \big) \bigg] \\
&\times \bigg[ \big(1+4e^2\big) \sin \omega \big( \alpha^k \gamma^l + \mathrm{sym} \big) + \big(1-e^2\big) \cos \omega \big( \beta^k \gamma^l + \mathrm{sym} \big)  \bigg] \; .
\end{split}
\end{align}
However, there seems to remain an additional dependence on $\omega$ which we do not expect. This spurious dependence can be removed completely by using the fact that $\mathcal{A}_{ij;kl}$ is antisymmetric under the exchange of the pairs $(i,j)$ and $(k,l)$. Taking the anti-symmetrization of the above expression, we are led to
\begin{equation}
\frac{\mu^2 a^4 (1-e^2)(1+4e^2)}{8} \left( (\alpha^i \gamma^j + \mathrm{sym}) (\gamma^k \beta^l +\mathrm{sym}) - (\gamma^i \beta^j + \mathrm{sym}) (\alpha^k \gamma^l +\mathrm{sym}) \right) \; ,
\end{equation}
so that all angular dependence drops out. Putting all together, we find the final expression for $\mathcal{B}_{ij;kl}$:
\begin{align}
\begin{split}
\mathcal{B}_{ij;kl} &= \frac{9 G_N^2 m_3^2 a^4}{32 \sqrt{G_N ma(1-e^2)}} \left[ -20 e^2(1-e^2)  (\alpha^i \beta^j +\mathrm{sym}) (4 \alpha^k \alpha^l - \beta^k \beta^l - \delta^{kl}) \right. \\
& \left. + (1-e^2)(1+4e^2) \left( (\alpha^i \gamma^j + \mathrm{sym}) (\beta^k \gamma^l +\mathrm{sym}) - (\beta^i \gamma^j + \mathrm{sym}) (\alpha^k \gamma^l +\mathrm{sym}) \right)  \right] \; ,
\end{split}
\end{align}
while the other tensor $\mathcal{A}_{ij;kl}$ is given by
\begin{align}
\mathcal{A}^{ij;kl} =& \frac{g_1(e_3)}{\sqrt{G_N M a_3^9}} \bigg[ \alpha_3^k \alpha_3^l \big( \alpha_3^i \beta_3^j + \mathrm{sym} \big) - \alpha_3^i \alpha_3^j \big( \alpha_3^k \beta_3^l + \mathrm{sym} \big)  \bigg]\\\nonumber& + \frac{g_2(e_3)}{\sqrt{G_N M a_3^9}} \bigg[ \beta_3^i \beta_3^j \big( \alpha_3^k \beta_3^l + \mathrm{sym} \big) - \beta_3^k \beta_3^l \big( \alpha_3^i \beta_3^j + \mathrm{sym} \big)  \bigg] \; ,
\end{align}
where the two dimensionless functions of the outer eccentricity $e_3$ are given by
\begin{align}
g_1(e_3) &= \frac{1}{48 e_3^2 (1-e_3^2)^{7/2}} \bigg[ 4 \big( 1 - \sqrt{1-e_3^2} \big) + e_3^2 \big( -8 + 9 \sqrt{1-e_3^2} \big) + e_3^4 \big(4 + 5\sqrt{1-e_3^2} \big) \bigg] \; , \\
g_2(e_3) &= \frac{1}{48 e_3^2 (1-e_3^2)^{5/2} } \bigg[ 4 \big( -1 + \sqrt{1-e_3^2} \big) + e_3^2 \big( 4 + \sqrt{1-e_3^2} \big)\bigg] \; .
\end{align}

Plugging the expressions of the basis vectors in terms of osculating angles, we have checked that our formula~\eqref{eq:L1Mean} exactly recovers the LPE with quadrupole-squared terms derived in~\cite{Will:2020tri}.

\section{From contact elements to orbital elements} \label{app:contact}

In this Appendix we explore the difference between contact elements and orbital elements, the two kinds of osculating elements that allow to describe the three-body system efficiently. As already remarked, the difference between these two sets of elements is of 1PN order, with the contact terms being particularly useful to repackage various PN corrections in the effective action. 
This analysis allows us to compare our results with some of the results of~\cite{PhysRevD.89.044043,Will:2014wsa} concerning the evolution of conserved quantities.

We will now write down explicitly the relation between these two sets of elements to 1PN order, following~\cite{1988CeMec..43..193R}. As discussed in Section \ref{sec:contact}, the key difference between the two sets of variables is the fact that the momentum is not simply proportional to the velocity in the PN expansion. Therefore it is useful to inspect the relation between $\mathbf{p}$ and $\mathbf{v}$ at 1PN:
\begin{equation}\label{eq:rel_p_v}
	\mathbf{p} = \mathbf{v} + \frac{\partial \mathcal{R}_\mathrm{1PN}}{\partial \mathbf{v}} = \mathbf{v} + \bigg( \frac{1-3\nu}{2} v^2 + \frac{G_N m(3+\nu)}{r} \bigg) \mathbf{v} + \frac{G_N m \nu}{r} (\mathbf{n} \cdot \mathbf{v} ) \mathbf{n} \; ,
\end{equation}
where we recall that in all Appendices we use the convention of dividing the Lagrangian by the reduced mass $\mu$, so that $\mathbf p$ and $\mathbf v$ have same dimension.
In the 1PN term we could use indifferently $\mathbf{v}$ or $\mathbf{p}$ since the difference would be of 2PN order.
The easiest orbital elements to relate to contact elements are the angles $\tilde{\Omega}$ and $\tilde{\iota}$. Indeed, since the momentum $\mathbf{p}$ is still in the plane of the motion, the definition of $\tilde{\Omega}$ and $\tilde{\iota}$ do not get affected by the difference between momentum and velocity. Thus,
\begin{align}
	\tilde \Omega &= \Omega \; , \\
	\tilde \iota &= \iota \; .
\end{align}
Let us start the non-trivial computations with the semimajor axis. We have the following definitions:
\begin{align}
	- \frac{G_N m}{2 \tilde a} &= \frac{v^2}{2} - \frac{G_N m}{r}\label{eq:def_a} \; , \\
	- \frac{G_N m}{2 a} &= \frac{p^2}{2} - \frac{G_N m}{r} \label{eq:def_tilde_a} \; .
\end{align}
Combining Eqs.\eqref{eq:rel_p_v}, \eqref{eq:def_a} and \eqref{eq:def_tilde_a} we get to $ \tilde a =  a + \Delta a$ with
\begin{equation}
	\Delta a = - \frac{2 G_N m}{(1 - \tilde e \cos \tilde \eta)^2} \bigg[ \frac{1}{2}(1-3\nu)(1+\tilde e \cos \tilde \eta)^2 + (3+\nu) (1+\tilde e \cos \tilde \eta) + \frac{\nu \tilde e^2 \sin^2 \tilde \eta}{1-\tilde e \cos \tilde \eta} \bigg] \; .
\end{equation}
In the above expression, using the contact or the osculating elements in the RHS makes no difference since we neglect terms of order 2PN ($\Delta a$ is a 1PN quantity). Let us now focus on the eccentricity. We have:
\begin{align}
	\vert \mathbf{r} \times \mathbf{v} \vert &= \sqrt{G_N m  \tilde a(1- \tilde e^2)} \; ,\\
	\vert \mathbf{r} \times \mathbf{p} \vert &= \sqrt{G_N m  a(1- e^2)} \; . \\
\end{align}
Splitting $\tilde e = e + \Delta e$ we get to
\begin{equation}
	\Delta e =  \frac{G_N m (1-\tilde e^2)}{\tilde a (1-\tilde e \cos \tilde \eta)^3} \bigg[ \cos \tilde \eta(1- \tilde e \cos \tilde \eta)(-7 + \nu -\tilde e (1-3\nu)\cos \tilde \eta) -\tilde e \nu \sin^2 \tilde \eta \bigg] \; .
\end{equation}
There finally remains to find the argument of perihelion $\omega$ and mean anomaly $u$. Let us denote by $f$ the true anomaly, representing the angle of the object along its trajectory on the ellipse, mesured from perihelion. Then one has $f + \omega=\tilde f + \tilde \omega$ since this corresponds to the true physical angle of the object, and should not depend on whether we use tilde quantities or not. Thus, $\Delta \omega = - \Delta f$.
To find $\Delta f$, one can use the definition of the radius vector:
\begin{equation}
	r = \frac{a(1-e^2)}{1+e\cos f} =  \frac{\tilde a(1-\tilde e^2)}{1+\tilde e\cos \tilde f} \; .
\end{equation}
This gives $\tilde f =  f + \Delta f$ with
\begin{equation}
	\Delta f = \frac{G_N m \sqrt{1-\tilde e^2} \sin \tilde \eta}{2 \tilde a \tilde e (1-\tilde e \cos \tilde \eta)^3} \bigg[14 -\tilde e^2 -2\nu +5 \tilde e^2 \nu -6\tilde e(2+\nu)\cos \tilde \eta - \tilde e^2(1-3\nu) \cos 2\tilde \eta \bigg] \; .
\end{equation}
Finally, to find $\Delta u$ (defined by $\tilde u =  u + \Delta u$)  one can use again the definition of $r=a(1-e \cos \eta)$ together with $\eta - e \sin \eta = u$ to find
\begin{align}
	\Delta u &= - \frac{G_N m}{8\tilde a \tilde e(1-\tilde e \cos \tilde \eta)^3} \bigg[ 2\big( -6\tilde e^4+\tilde e^2(1-15\nu)-4(7-\nu) \big) \sin \tilde \eta \nonumber \\
	&+ 2\tilde e \big(6(2+\nu) +2\tilde e^2(7+2\nu)-\tilde e^4(1-3\nu) \big)\sin 2\tilde \eta \nonumber\\
	&+ 2\tilde e^2\big(1-3\nu-\tilde e^2(6+4\nu)\big)\sin 3\tilde \eta - \tilde e^5(1-3\nu) \sin4\tilde \eta \bigg]\;.
\end{align}
To sum up our results, here are all the modifications to the osculating elements:
\begin{align}
	\Delta a &= - \frac{2 G_N m}{(1 - \tilde e \cos \tilde \eta)^2} \bigg[ \frac{1}{2}(1-3\nu)(1+\tilde e \cos \tilde \eta)^2 + (3+\nu) (1+\tilde e \cos \tilde \eta) + \frac{\nu \tilde e^2 \sin^2 \tilde \eta}{1-\tilde e \cos \tilde \eta} \bigg] \; , \label{eq:deltaa} \\
	\Delta e &=  \frac{G_N m (1-\tilde e^2)}{\tilde a (1-\tilde e \cos \tilde \eta)^3} \bigg[ \cos \tilde \eta(1- \tilde e \cos \tilde \eta)(-7 + \nu -\tilde e (1-3\nu)\cos \tilde \eta) -\tilde e \nu \sin^2 \tilde \eta \bigg] \; , \\
	\Delta \omega &= - \frac{G_N m \sqrt{1-\tilde e^2} \sin \tilde \eta}{2 \tilde a \tilde e (1-\tilde e \cos \tilde \eta)^3} \bigg[14 -\tilde e^2 -2\nu +5 \tilde e^2 \nu -6\tilde e(2+\nu)\cos \tilde \eta - \tilde e^2(1-3\nu) \cos 2\tilde \eta \bigg] \; , \\
	\Delta u &= - \frac{G_N m}{8\tilde a \tilde e(1-\tilde e \cos \tilde \eta)^3} \bigg[ 2\big( -6\tilde e^4+\tilde e^2(1-15\nu)-4(7-\nu) \big) \sin \tilde \eta \nonumber \\
	&+ 2\tilde e \big(6(2+\nu) +2\tilde e^2(7+2\nu)-\tilde e^4(1-3\nu) \big)\sin 2\tilde \eta \nonumber\\
	&+ 2\tilde e^2\big(1-3\nu-\tilde e^2(6+4\nu)\big)\sin 3\tilde \eta - \tilde e^5(1-3\nu) \sin4\tilde \eta \bigg] \; , \\
	\Delta \Omega &= \Delta \iota = 0 \; , \label{eq:deltaiota} \\
	\Delta \bm \alpha &= \bm \beta \Delta \omega \; , \quad \Delta \bm \beta = - \bm \alpha \Delta \omega \; .
\end{align}

These formulas relate the instantaneous values of these two sets of elements. However, it is more interesting to analyse the difference in the orbit-averaged elements, denoted with an $\hphantom{}_L$ subscript. Using our previous splitting between long-timescale and short-timescale variables (see Appendix~\ref{app:beyond_adiab}), one can write e.g for the eccentricity $\tilde e =  e + \Delta e =  e_L + e_S + \Delta e $ with $\langle e_S \rangle =0$. Thus, we can identify:
\begin{equation}
	\tilde e_L =  e_L + \langle \Delta e \rangle \; , \quad \tilde e_S =   e_S + \Delta e - \langle \Delta e \rangle \; ,
\end{equation}
and similarly for the other osculating elements. This gives the final relation for the orbit-averaged contact and orbital elements:

\begin{mybox}
	\begin{align} \label{eq:shift_contact_osculating}
		\begin{split}
			\tilde a_L &=  a_L +  G_N m \bigg[  9 - \frac{16}{\sqrt{1-\tilde e_L^2}} - \nu \bigg(5 - \frac{6}{\sqrt{1-\tilde e_L^2}} \bigg) \bigg]  \; , \\
			\tilde e_L &=  e_L  - \frac{G_N m \tilde e_L}{\tilde a_L} (8 -3\nu) \frac{\sqrt{1-\tilde e_L^2}}{1+\sqrt{1-\tilde e_L^2}} \; , \\
			\tilde\omega_L &=  \omega_L \; , \quad  \tilde \Omega_L =  \Omega_L \; , \quad \tilde \iota_L =  \iota_L \; , \quad \tilde u_L =  u_L \; .
		\end{split}
	\end{align}
\end{mybox}

We can now describe the main physical effects stemming from this post-Newtonian shift of the osculating elements:

\begin{itemize}
	\item The first and most important point to notice is that the shifts written in Eq.~\eqref{eq:shift_contact_osculating} will stay small at \textit{any} moment in time. In other words, they cannot accumulate a small effect over long timescales to get an important effect\footnote{This is, as long as the elements are computed within the time interval in which our effective field theory is valid.}, like it happens for quadrupolar and post-Newtonian perturbations in the LPE; rather, the two sets of osculating elements will differ by a quantity which is of post-Newtonian order at any time. Thus, replacing contact elements by orbital ones will not make a qualitative difference concerning the long-timescale evolution of the binary system.

	\item Plugging the shifts $\tilde e_L =  e_L - \langle \Delta e \rangle$ and $\tilde a_L =  a_L - \langle \Delta a \rangle$ implied by Eq.~\eqref{eq:shift_contact_osculating} in the kinetic term of the Lagrangian~\eqref{eq:def_R} shifts the canonical momenta $G$ and $H$ as
	\begin{equation}
		\frac{\Delta G}{\tilde G} = \frac{\Delta H}{\tilde H} =  \frac{e \langle \Delta e\rangle}{1-\tilde e^2} - \frac{\langle \Delta a \rangle }{2\tilde a} = \frac{G_N m}{2\tilde a} (7-\nu) \; .
	\end{equation}
	In other words, this replaces the effective Newtonian spin of the binary system $\mathbf{J} = \mu \sqrt{G_N m \tilde a(1-\tilde e^2)} \bm \gamma$ with its 1PN counterpart given by
	\begin{equation} \label{eq:J1PN}
		\mathbf{J}_\mathrm{1PN} =  \mu \sqrt{G_N m  a(1-e^2)} \tilde{\bm \gamma} = \mu \sqrt{G_N m \tilde a(1-\tilde e^2)} \bm \gamma \bigg(1+\frac{G_N m}{2\tilde a} (7-\nu) \bigg) \; .
	\end{equation}
	On the other hand, the above expression corresponds to the 1PN expression of the conserved angular momentum that one can find in e.g.~\cite{AIHPA_1985__43_1_107_0}, averaged over one orbit of the binary system. This is a nontrivial check of the validity of our computation.

	\item The most interesting effect coming from these shifts of $a$ and $e$ is that the semimajor axis $\tilde a$ is not conserved in time. Indeed, it is the contact element $ a$ which is conserved in time, but it is related to the osculating $\tilde a$ with a formula involving the eccentricity, eq.~\eqref{eq:shift_contact_osculating}. Since the eccentricity is itself allowed to vary, one should have a variation of $\tilde a$ over time at 1PN order.
	This effect has already been discussed in~\cite{PhysRevD.89.044043,Will:2014wsa} where it was derived using the 1PN equations of motion. However, our treatment makes clear the fact that such an effect cannot accumulate over a long timescale and give appreciable variation in the semimajor axis $a$ as discussed before.
	
	On the other hand, a detailed comparison shows that our formula for the variation of $\tilde a$ and the one given in~\cite{PhysRevD.89.044043,Will:2014wsa} seem to be in disagreement. The source of this apparent incompatibility can be traced back to the fact that our averaging procedure is somewhat different than the one discussed in~\cite{PhysRevD.89.044043,Will:2014wsa}. Indeed, the following averages differ at 1PN quadrupolar order:
	\begin{equation}
		\frac{\mathrm{d} \langle a \rangle}{\mathrm{d} t} \neq \left \langle \frac{\mathrm{d} a}{\mathrm{d} t} \right \rangle \; .
	\end{equation}
	The LHS of this equation corresponds to the quantity that we compute in this article. On the other hand, the variation computed in~\cite{PhysRevD.89.044043,Will:2014wsa} corresponds to the RHS of this equation, which by definition is equal to $(a(T)-a(0))/T$. Thus, the results of~\cite{PhysRevD.89.044043,Will:2014wsa} concern the evolution of the \textit{initial value} of osculating elements after one inner period, while we are interested in the evolution of the \textit{mean value} of osculating elements. This difference also shows itself in the conservation of energy and angular momentum: while the work in~\cite{PhysRevD.89.044043,Will:2014wsa} proves exact conservation of \textit{initial} PN energy and angular momentum, we are able to prove conservation of an \textit{averaged} PN energy and angular momentum.
	
\end{itemize}

\bibliographystyle{unsrt}
\bibliography{Kozai}

\end{document}